\documentclass[aps,prb,twocolumn,superscriptaddress,floatfix,showpacs]{revtex4-1}
\usepackage{tikz}
\usepackage{amsmath}
\usepackage{makeidx,amssymb,amsmath,amsthm,graphicx}
\usepackage[toc,page]{appendix}
\usepackage{xcolor}

\begin{document}


\title{
GdRh$_2$Si$_2$: An exemplary tetragonal system for antiferromagnetic
order with weak in-plane anisotropy}

\author{K.Kliemt}
\email[]{kliemt@physik.uni-frankfurt.de}
\affiliation{Physikalisches Institut, Goethe-Universit\"at Frankfurt/M, 60438 Frankfurt/M, Germany}
\author{M.Hofmann-Kliemt}
\affiliation{Fachbereich Mathematik, Technische Universit\"at Darmstadt, 64289 Darmstadt, Germany}
\author{K.Kummer}
\affiliation{European Synchrotron Radiation Facility, 71 Avenue des Martyrs, 38043 Grenoble, France}
\author{F.Yakhou-Harris}
\affiliation{European Synchrotron Radiation Facility, 71 Avenue des Martyrs, 38043 Grenoble, France}
\author{C.Krellner}
\affiliation{Physikalisches Institut, Goethe-Universit\"at Frankfurt/M, 60438 Frankfurt/M, Germany}
\author{C.Geibel}
\affiliation{Max Planck Institute for Chemical Physics of Solids, 01187 Dresden, Germany}
\date{\today}

\begin{abstract}
The anisotropy of magnetic properties commonly is introduced in textbooks using
the case of an antiferromagnetic system with Ising type anisotropy. 
This model presents huge anisotropic magnetization and a pronounced 
metamagnetic transition and is well-known and well-documented 
both, in experiments and theory. In contrast, the case of an 
antiferromagnetic  $X$-$Y$ system with weak in-plane anisotropy
is only poorly documented. We studied the anisotropic magnetization 
of the compound GdRh$_2$Si$_2$ and found that it is a 
perfect model system for such a weak-anisotropy setting because the Gd$^{3+}$ ions 
in GdRh$_2$Si$_2$ have a pure spin moment of S=7/2 which orders in a simple 
AFM structure with ${\bf Q} = (001)$.
We observed experimentally in $M(B)$ a 
continuous spin-flop transition and domain effects for field applied 
along the $[100]$- and the $[110]$-direction, respectively. 
We applied a mean field model for the free energy to describe our data and combine
it with an Ising chain model to account for domain effects. 
Our calculations reproduce the experimental data very well.
In addition, we performed magnetic X-ray scattering 
and X-ray magnetic circular dichroism measurements, 
which confirm the AFM propagation vector to be ${\bf Q} = (001)$
and indicate the absence of polarization 
on the rhodium atoms.

\end{abstract}

\maketitle

\def\neel{{N\'eel} }
\def\FA{F_{\rm an}}
\def\CA{C_{\rm an}}
\def\MS{M_{\rm sat}}
\def\EDF{E_{\rm df}  }
\def\text#1{{\rm #1}}
\def\i{\item}
\def\[{\begin{eqnarray*}}
\def\]{\end{eqnarray*}}
\def\bv{\begin{verbatim}}
\def\ev{\end{verbatim}}
\def\ganz{Z}
\def\3{\ss}
\def\reel{{\cal}R}
\def\platz{\;\;\;\;}
\def\beginvector{\left(\begin{array}{c}   }
\def\endvector{\end{array}\right)}
\def\fff{\frac{3}{k_B} F }
\def\vec#1{ {\rm \bf #1  } }
\def\KBMUEF{\frac{3k_B}{\mu_{\rm eff}^2 }}


\section{Introduction}

Magnetic compounds usually present a clear anisotropy
respective to the direction of an applied magnetic field or the direction of 
the measurement. It origins on one hand from on-site interactions 
like for instance crystal field effects, 
which result in an orientation dependence
of the size of magnetic moments. On the other hand it stems from the inter-site interactions
which may depend on the orientation of the interacting moments 
due to for example spin-orbit coupling. This results in different 
values of the magnetic susceptibility along different directions,
but also in specific field-dependent effects like e.g. metamagnetic transitions,
where either the slope of the magnetization or the magnetization
itself show a step-like behaviour as a function of field.
In textbooks on solid state magnetism,  anisotropy is often introduced 
and discussed in the context of an axial Ising system, 
where it leads to a first order metamagnetic transition 
for a field applied along the easy axis \cite{Blundell2011}. 
This Ising case is well known and well documented 
both, experimentally and theoretically \cite{Wolf2000}. 
In contrast, the in-plane anisotropy of an easy-plane system 
has been much less studied and is poorly documented, 
despite its relevance for quite a number of compounds. 
Potential candidates are compounds with global and local
(at the magnetic site) tetragonal or hexagonal symmetry and therefore all compounds
 crystallizing 
in the tetragonal ThCr$_2$Si$_2$ structure type, 
which is one of the most common among intermetallic compounds. 
Some of them like ErRu$_2$Ge$_2$ and 
the RNi$_2$B$_2$C (R = Ho, Dy, Er) systems 
present a huge in-plane anisotropy which results 
in complex metamagnetic processes \cite{Suzuki2004,Canfield1997,Amici1998}. \\
Here we study the opposite case, 
namely the effect of a small in-plane anisotropy. 
GdRh$_2$Si$_2$ is a layered antiferromagnet and crystallizes in the tetragonal ThCr$_2$Si$_2$ structure \cite{Felner1984}. 
Its \neel temperature $T_{\rm N}\approx 107\,\rm K$ \cite{Czjzek1989, Kliemt2015} is the highest in the RRh$_2$Si$_2$ (R = rare earth) series. 
For the half-filled 4f shell of the Gd$^{3+}$ ions the total orbital momentum L vanishes 
and the total angular momentum is equal to the spin momentum J=S=7/2. 
Studies of Gd compounds are of special interest since the L=0 configuration leads to the absence of spin-orbit coupling 
and to the insensitivity of the $^8\rm S_{7/2}$ ground state to crystalline electric field (CEF) effects. 
In this compound, a slight enhancement of the effective magnetic moment $\mu_{\rm eff}=8.28\,\mu_{\rm B}$ \cite{Czjzek1989, Kliemt2015} 
and the inplane ordering of the magnetic moments \cite{Felner1984} was observed.
In the course of our investigation of GdRh$_2$Si$_2$, we observed weak, 
but very clear features in the magnetization at low fields.
Analyzing these data, we found 
that this compound presents an exemplary case 
for a simple antiferromagnetic (AFM) order with in-plane ordered moments
and weak in-plane anisotropy. 
We set up a magnetic mean field model which we combined with an Ising chain model
to account for domain effects, which perfectly reproduces 
the experimental data. 
Therefore, our study provides a firm basis for analyzing similar systems which has not been reported so far. 
The determination of the magnetic structure of Gd-compounds by neutron scattering is hindered due to the large absorption cross-section of Gd. We hereby present a different approach. 
Additional magnetic X-ray diffraction and X-ray magnetic circular dichroism (XMCD) results indicate 
an AFM propagation vector $\vec{Q} = (001)$, 
the same as in all RRh$_2$Si$_2$ compounds 
with a stable trivalent rare earth R \cite{Slaski1983, Szytula1984, Melamud1984, Quezel1984, Szytula1989},
and the absence of a measurable polarization on Rh. 
Recently, the surface magnetism of GdRh$_2$Si$_2$ was subject of a thorough investigation by angle-resolved photoemission spectroscopy (ARPES) 
which yield a large spin-splitting of the surface states \cite{Guettler2016}.
To clearly separate surface and bulk-related effects, the knowledge of the bulk magnetization provided by our study is indispensible. 
The paper is organized as follows: In Sec.~\ref{exp}, the experimental details are summarized. The susceptibility, discussed in Sec.~\ref{sus}, indicates the ordering of the magnetic moments along the $[110]$-direction in the basal plane of the tetragonal lattice. X-ray magnetic scattering yields an A-type magnetic structure as described in Sec.~\ref{App1}.
XMCD results, presented in Sec.~\ref{App2}, yield the absence of a contribution of Rh to the magnetization. A mean-field model was developed for an A-type magnetic structure, Sec.~\ref{mean}, which fits the magnetization data very well and hereby confirms the assumed direction of inplane ordering.


\section{Experiment}\label{exp}
The details of the growth procedure 
of GdRh$_2$Si$_2$ single crystals were reported recently \cite{Kliemt2015}. 
Magnetization measurements  
were performed using the Vibrating Sample Magnetometer (VSM) option of
a Quantum Design Physical Properties Measurement System.   
The X-ray absorption measurements were performed at the ID32 beamline 
of the European Synchrotron Radiation Facility. 
The sample was cleaved in ultra-high vacuum at a base pressure 
of 2$\cdot 10^{-10}$\,mbar and a temperature of $T = 5\rm\,K$. 
A field of 9\,T was applied parallel to the $[001]$-direction 
of the sample which was aligned to the $\vec{k}$ vector of the 
incoming X-ray beam ($\vec{k}\parallel\vec{B}\parallel$ 001). 
Then the X-ray magnetic circular dichroism (XMCD), i.e. 
the difference in the absorption of circular left 
and circular right polarized light was measured by scanning
the photon energy over the Gd M$_{4,5}$ and Rh M$_{2,3}$ 
absorption edges and detecting the total electron yield signal.
The measurements performed on three different samples gave the same results.
The soft X-ray resonant magnetic scattering measurements 
have also been performed at the ID32 beamline  
using the four-circle goniometer installed in the inelastic
X-ray scattering spectrometer as a diffractometer. 
The $(002)$ reflection and a photon energy of 1.4\,keV 
was used for sample alignment. The magnetic scattering 
was done at the Gd M$_5$ edge at 1.185\,keV, 
resonant to the Gd 3d$_{5/2}$ $\to$ 4f excitations. 

\begin{figure}
\centering
\includegraphics[width=0.5\textwidth]{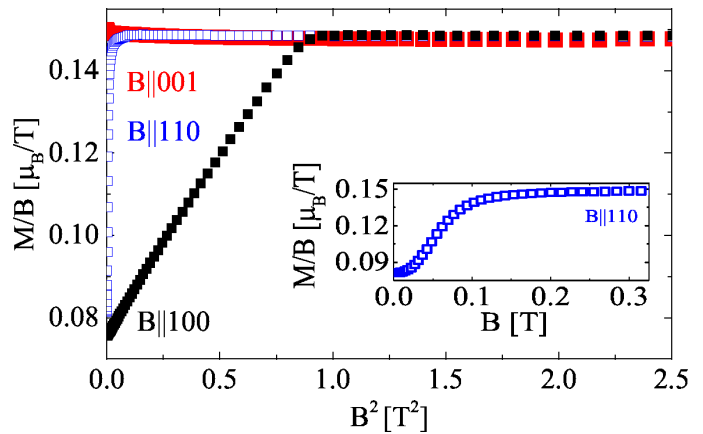}
\caption[]{
Observed anisotropy in the low field susceptibility per Gd-ion $M/B$ at $T$=3.5\,K.
The red squares for $\vec{B}\parallel 001$ show a constant susceptibility.
The blue open squares for $\vec{B}\parallel 110$ show the domain flip. 
The enlarged view for lower fields is shown in the inset.
The black squares for $\vec{B}\parallel100$ show a metamagnetic 
spin-flop transition.
}
\label{observ}
\end{figure}

\begin{figure}
\centering
\includegraphics[width=0.5\textwidth]{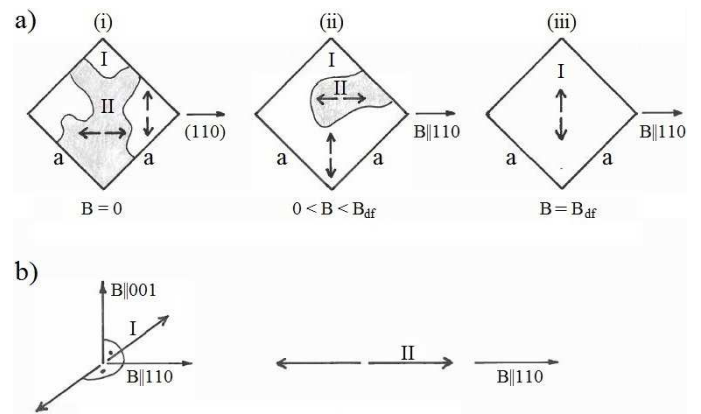}
\caption[]{a) Schematic diagram of the domain distribution upon increasing field for
 $\vec{B}\parallel 110$; b) {\it Left:} For $\vec{B}\parallel 110$ and $\vec{B}\parallel 001$, 
the field is applied perpendicular to the sublattice magnetization of domain I; {\it Right:} For $\vec{B}\parallel 110$, 
the field is applied parallel to the sublattice magnetization of domain II.
}
\label{observ9}
\end{figure}

\section{Experimental results and discussion}\label{sus}
\begin{figure}
\centering
\includegraphics[width=.5\textwidth]{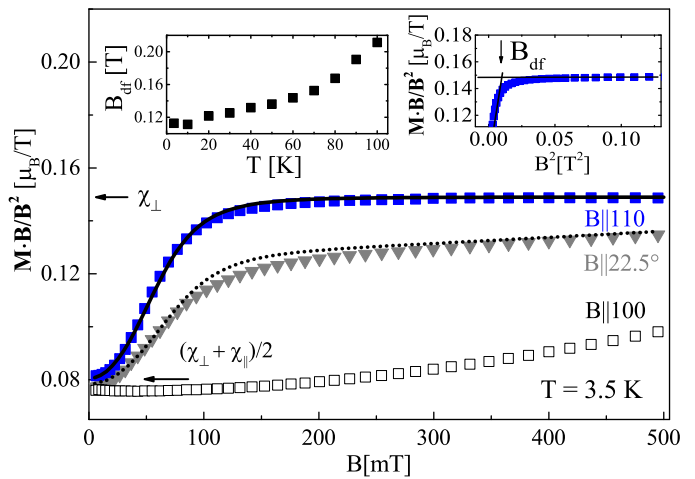}
\caption[]
{Field dependence of the inplane susceptibility per Gd-ion. 
The measured curve for $\vec{B}$ parallel 
to the $[110]$-direction (blue closed squares) 
can be fitted (solid line) with Eqn.~(\ref{formula110}). 
The left inset shows the temperature dependence of $B_{\rm df}$.
This field can be obtained at the crossing point of the two linear fits
as shown in the right inset. 
\label{fit110bild}
For $\vec{B}\parallel$22.5$^{\circ}$, there is a slight
deviation between 100 and 200\,mT
of the measured curve (grey triangles)
and the prediction of the one dimensional Ising chain model
(Eqn.~(\ref{22fit}), dotted line).
This deviation occurs in the same fashion but smaller
in the $[110]$-direction. Since for $\vec{B}\parallel$22.5$^{\circ}$, 
a direction which is not a main symmetry direction, the 
torque is non-zero, we label the ordinate $\vec{M}\cdot\vec{B}/\vec{B}^2$ instead of $M/B$.
}
\label{22fitbild}
\end{figure}

\begin{figure}
\centering
\includegraphics[width=0.5\textwidth]{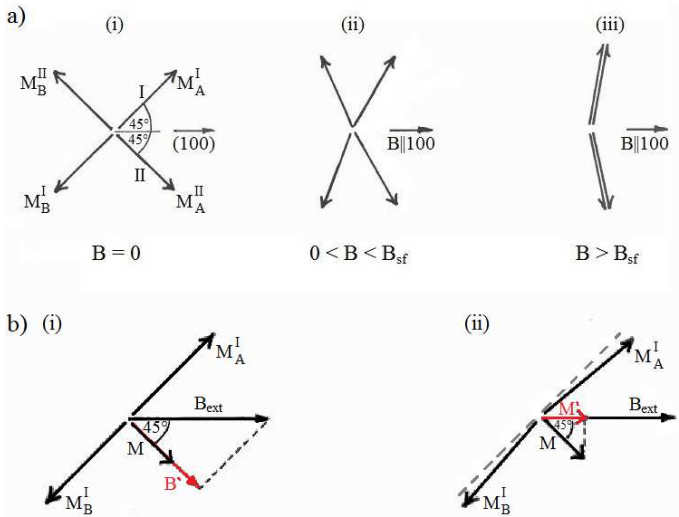}
\caption[]{ a) Schematic diagram of the reorientation of the moments ("spin-flop") upon increasing field for
 $\vec{B}\parallel 100$; b) (i) The acting field is $B^{\prime}=1/\sqrt{2} B_{\rm ext}$ due to the projection onto the direction of the magnetization $\vec{M}$; (ii) The magnetization is reduced due the projection onto the direction of the external field $M^{\prime}=1/\sqrt{2} M$.}
\label{observ10}
\end{figure}

\begin{figure}
\centering
\includegraphics[width=.5\textwidth]{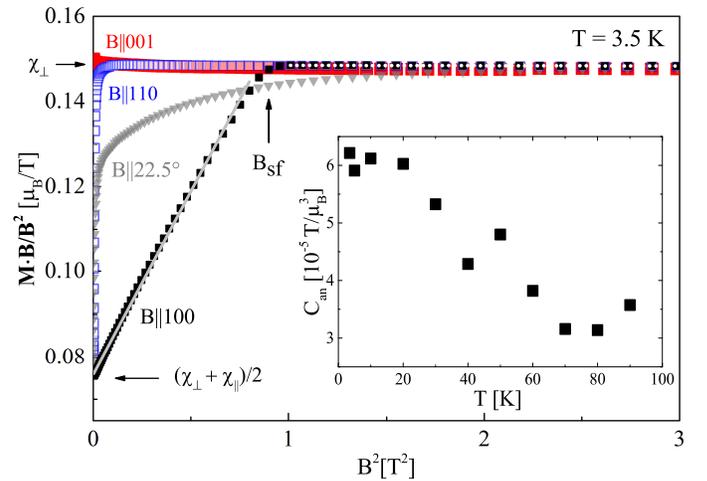}
\caption[]{
Susceptibility per Gd-ion $\vec{M}\cdot\vec{B}/\vec{B}^2$ versus $B^2$ at $T=3.5$\,K for 
$\vec{B}$ parallel to different crystal orientations. 
The model predicts that the data recorded with 
the field along the main symmetry directions run into the plateau exactly, 
while the 22.5$^{\circ}$ curve reaches the plateau asymptotically. 
For low (grey solid line) and higher fields (white dotted line)
parallel to the $[100]$-direction, 
Eqn.~(\ref{Dir100Fit}) fits the data quite well. 
In the inset, the temperature dependence of the anisotropic term,
$\CA(T)$, is shown. 
$\CA(T)$ was calculated from the spin-flop field 
$B_{\rm sf}$ which was determined from the change in 
the slope of $M(B)$ measured at different temperatures below 100\,K. 
}
\label{fit100}
\end{figure}

For an uniaxial Ising system a spin-flop transition
for field along the easy direction and a continuous increase 
of the magnetization for transversal field is predicted and observed. 
The situation is different in our case where we have a system 
with a fourfold symmetry and the moments being aligned 
in the basal plane at 4.1\,K \cite{Felner1984}. 
Our data, presented in this section, indicate the alignment along the $[110]$-direction. We now demonstrate how this assumption was deduced from the observed susceptibility. 
The susceptibility $\chi(B)=M/B$ curves in Fig.~\ref{observ}
show the presence of an inplane anisotropy in 
this system within the ordered phase. 
For $\vec{B} \parallel 001$, the susceptibility 
$M/B=0.149\,\mu_B/T$ in the ordered phase is constant
as a function of the magnetic field,
while for $\vec{B}$ parallel to $[110]$ or $[100]$ 
the susceptibility changes strongly below 1T. 
In the limit $\vec{B}=0$, the susceptibility along the $[100]$- and along the $[110]$-direction is identical within the experimental accuracy, but presents half the value of the susceptibility for
field along the $[001]$-direction. Upon increasing field, the susceptibility increases for 
both in-plane directions, but in a very different way and on different field scales. We now dicuss two particular cases in detail. \\
In the first case, $\vec{B}\parallel 110$, the susceptibility 
shows a shaped increase with $\vec{B}$, which starts at a field of merely 20 mT and smoothly saturates at the same value as for $\vec{B}\parallel 100$ for a field of the order of 250 mT 
at 3.5\,K (see inset of Fig.~\ref{observ}).
We explain the observed behaviour by domain effects (Fig.~\ref{observ9}a).
In our system with the fourfold symmetry and the moments being aligned in the basal plane
there must exist two domains namely
domain I with the moments perpendicular to the field and  
domain II with the moments aligned parallel to the field. 
The bulk contains two regions with relative volume sizes $a^I, a^{II}$ 
corresponding to the two possible alignments of the magnetic moments.
Since we do not observe any hystereses effects the distribution of the domains is that of the thermodynamic equilibrium. 
At $\vec{B}=0$, both domains have the same energy and thus the same probability and the same volume sizes (Fig.~\ref{observ9}a(i)).
With $\vec{B}\parallel 110$, the field is applied perpendicular to the sublattice magnetization of domain I,
but parallel to the sublattice magnetization  
of domain II (Fig.~\ref{observ9}b).
In the limit $\vec{B}=0$, the susceptibility of domain I is identical to that 
for field along $[001]$, since in both cases the field is aligned transversal 
to the orientation of the ordered moments. In contrast, the susceptibility 
of domain II is almost zero, 
as for the Ising case in longitudinal field. When increasing $\vec{B}$, domain I gains polarization energy in the external field, while for domain II the gain is negligible. Therefore domain I becomes energetically more favourable, and the size of domain II shrinks in favour of domain I (Fig.~\ref{observ9}a(ii)). 
This results in an increase in $\chi(B)$ up to the same value as observed 
for field along $[001]$. We call this case a domain flip. The susceptibility $\chi(B)$ can be 
quantitavely described as shown in Fig.~\ref{22fitbild}.\\
In the second case, $\vec{B}\parallel 100$ (Fig.~\ref{observ}, black squares), we observe a linear increase of $\chi(B)$, 
from half the value as for 
$\vec{B}\parallel 001$ until a sharp saturation at the same value as for $\vec{B}\parallel 001$ (blue and red symbols in Fig.~\ref{observ}).
This can also easily be explained. 
We assume that the susceptibility for field aligned parallel to the sublattice magnetization 
$\chi_{\parallel}$ 
is negligible and from the fact that the moments are aligned in the basal plane we know that the susceptibility for $\vec{B}\parallel 001$ is $\chi_{\perp}$.
For $\vec{B}\parallel 100$ and in the limit $\vec{B}=0$, 
the easy axis of both domains makes an angle of 45$^\circ$ with the 
direction of the external magnetic field (Fig.~\ref{observ10}a(i)). Thus, the external field component perpendicular to the
ordered moment amounts to only $1/\sqrt{2}$ of the total value of the external field (Fig.~\ref{observ10}b(i)). 
Tilting of the AFM ordered moments towards this perpendicular direction leads to a magnetization which makes an angle of 45$^\circ$ to the external field direction. Thus, projecting this magnetization on the 
direction of the external field leads to a further factor of $1/\sqrt{2}$ (Fig.~\ref{observ10}b(ii)). Therefore, the susceptibility for $\vec{B}\parallel 100$ is reduced by a factor of 2 in total compared to $\chi_{\perp}$. The increase of the magnetic field leads to
a continuous reorientation of the ordered moments perpendicular to the external field (Fig.~\ref{observ10}a(ii),(iii)), 
and therefore to the recovery of the full susceptibility in a transversal field. In this case, both domains behave in an identical way. 
This rotation of local spin directions is called 
a continuous spin-flop transition \cite{Stryjewski1977} and can be quantitatively described as shown in Fig.~\ref{fit100}. \\
Our magnetization data suggest the alignment of the moments along the 
$[110]$-direction in GdRh$_2$Si$_2$. 
To verify this assumption, we set up a mean field description 
to model the magnetization measurements.


\section{X-ray magnetic scattering}
\label{App1}
\begin{figure}
    \centering
\includegraphics[width=.5\textwidth]{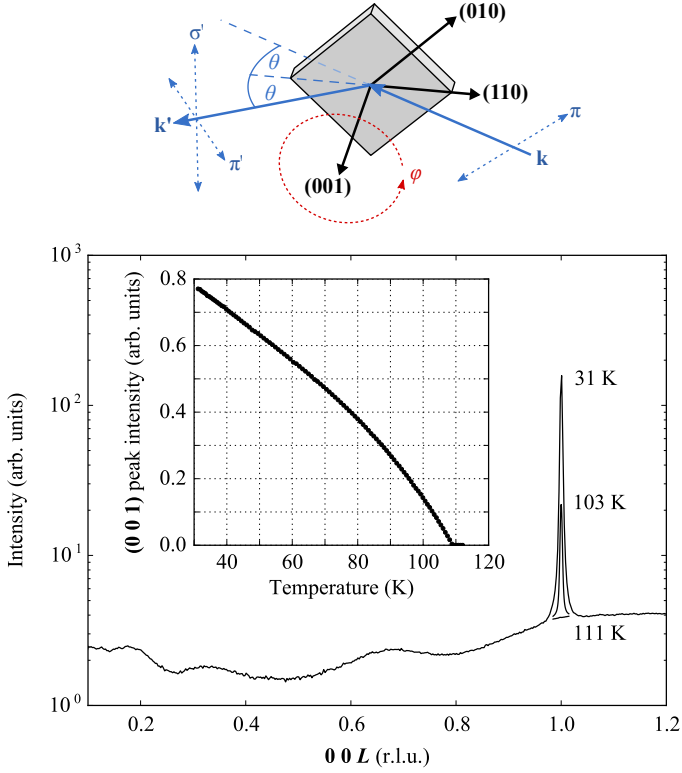}
    \caption{ X-ray magnetic scattering: Scattering geometry (top) and $(0\,0\,L)$ scans at different temperatures using $\pi$ polarized light (bottom). Below $T_N$ a magnetic Bragg peak is observed at the charge-forbidden $(0\,0\,1)$ position. The inset shows the intensity of the $(0\,0\,1)$ peak vs. temperature and is consistent with the previously reported $T_N = 107\,$K of GdRh$_2$Si$_2$.}
\label{Kurt2}
 \end{figure}

Soft X-ray resonant magnetic scattering in zero field confirms the assumption 
of ferromagnetic planes stacked antiferromagnetically in c-direction. 
As shown in Fig.~\ref{Kurt2} a magnetic reflection arises below 
$T_{\rm N}$ at $\vec{Q}=(0\,0\,1)$ which is a charge-forbidden reflection 
of the tetragonal lattice. In the AFM phase the two Gd atoms in 
the unit cell become magnetically non-equivalent which gives rise 
to anomalous X-ray scattering at the charge forbidden reflections $(0\,0\,L)$, 
$L=2n+1$. No further reflections are observed at non-integer values of $L$
which confirms a simple commensurate order with 
an antiparallel alignment of adjacent Gd layers. 
The evolution of the intensity of the $(0\,0\,1)$ peak with temperature 
is shown as an inset. Because of its magnetic nature the intensity 
of the $(0\,0\,1)$ reflection can be directly 
related to the magnetic properties of GdRh$_2$Si$_2$.
Note that in the chosen scattering geometry and using $\pi$ polarised light the intensity of the $(0\,0\,1)$ reflection does only depend on the absolute value of the magnetic moment in the basal plane but not its orientation (see App.~\ref{app:magnetic_scattering}). Therefore, the observed temperature dependence of the $(0\,0\,1)$ reflection is independent of how many domains are probed and how they are oriented.

\section{X-ray magnetic circular dichroism}
\label{App2}
\begin{figure}
\centering
\includegraphics[width=.5\textwidth]{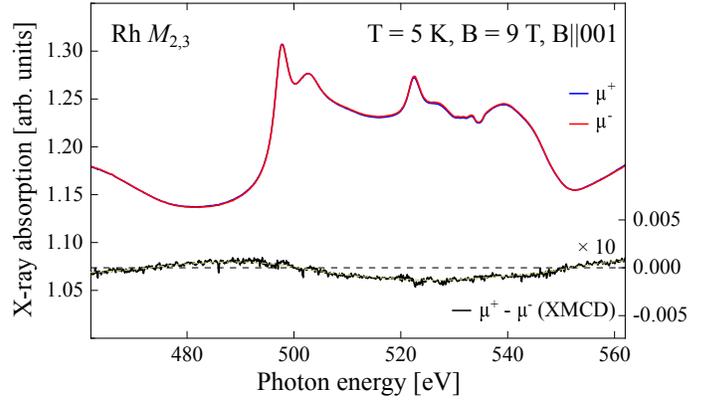}
\caption[]{XMCD signal at the Rh M$_{2,3}$ absorption edge. 
No XMCD signal is detected at the Rh sites which additionally 
rules out a significant Rh contribution to the total magnetic moment.
}
\label{Kurt}
\end{figure}

Utilizing Hund's rules, which usually hold well 
for the description of localized 4f electron states, 
an effective magnetic moment 
of $\mu_{\rm eff}=7.94\,\mu_B$ is predicted. Experimentally, 
a larger value of $(8.28 \pm 0.10)\,\mu_B$
was determined \cite{Kliemt2015}. In order to evaluate 
a possible contribution of Rh sites to the total 
magnetic moment we performed 
X-ray magnetic circular dicroism (XMCD) measurements 
at the Gd M$_{4,5}$ and Rh M$_{2,3}$ edges, respectively, 
which allows to separate the Gd and Rh contributions. 
The detected XMCD signal is shown in Fig.~\ref{Kurt} for $B=9\,$T 
and $T=5\,$K. A large XMCD signal is detected at the Gd M$_{4,5}$ 
edges (not shown). We have evaluated the Gd spin and orbital moment 
aligned along the field direction $M_S=(1.35 \pm 0.05)\,\mu_B$, 
$M_L=(0.00 \pm 0.03)\,\mu_B$  using the XMCD sum rules 
\cite{Carra1993, Krishnamurthy2009} and obtained values
which agree well with the total moment obtained from
our magnetization measurements and with the expected
zero orbital moment of the Gd 4f$^{7}$ configuration.
In contrast, no XMCD signal is detected at the Rh M$_{2,3}$ edges,
at least within the sensitivity of the experiment.
With the signal-to-noise ratio of our experiment, XMCD signals smaller than 0.2\% can be detected, corresponding to an aligned Rh moment of about 0.015 $\mu_B$ \cite{Kummer2016}, \cite{Sessi2010}. 
Assuming that at 5\,K and 9\,T only about 20\% of the total Rh moment is saturated, i.e. the same fraction as observed for Gd, this would put an upper limit of about 0.1 $\mu_B$ on the total magnetic moment per Rh site. 
Hence, our experimental XMCD data indicate that the too high 
effective magnetic moment compared to the theoretical
prediction cannot or, at least, not only be explained
by an additional magnetic contribution from the Rh ions.


\section{Mean field model}\label{mean}

\begin{figure}
\centering
\includegraphics[width=0.23\textwidth]{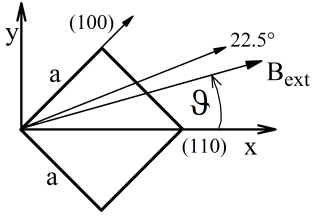}
\caption[]{
Choice of the coordinate system in the basal plane.
}
\label{coord}
\end{figure}

The results of our low-field magnetization measurement support the assumption that the magnetic moments are aligned along the $[110]$-direction at low temperature. 
We therefore choose the coordinate 
system such that the $x$-axis points in $[110]$-direction
and the  $y$-axis in $[\bar{1}10]$-direction 
of the tetragonal lattice as shown in Fig.~\ref{coord}. 
The main contribution to the magnetization in GdRh$_2$Si$_2$ 
comes from the Gd$^{3+}$-ions.
We utilize the result of our X-ray magnetic scattering result, that these magnetic 4f spins constitute FM layers 
with a simple alternating AFM staggering in c-direction.
The Gd$^{3+}$-ions of every second layer are
ensembled to one FM sublattice A
with the mean magnetization $\vec{M}_A$,
the intervening layers to a second FM sublattice B with $\vec{M}_B$.
The mean value  of the magnetization per ion in the whole bulk is 
$\vec{M}:=(\vec{M}_A + \vec{M}_B)/2$. We restrict the magnetization to two dimensions and 
model the magnetization measurement using the free energy
\begin{eqnarray}
F(\vec{M}_A,\vec{M}_B)&=&
-T\;S - \vec{M} \cdot\vec{B}+ \Phi(\vec{M}_A,\vec{M}_B)
\label{masterfree}
\end{eqnarray}
with the entropy $S$, the temperature $T$, the Zeeman term
$\vec{M} \cdot\vec{B}$ and a magnetic interaction energy
$\Phi(\vec{M}_A,\vec{M}_B)$.
The magnetizations $\vec{M}_A$ and $\vec{M}_B$
act as meanfield parameters,
describing the thermodynamic equilibrium, 
once the free energy is minimized.
The minimum of $F$ is not unique and
several domains exist, whose relative volume sizes
$a^{I}$, $a^{II}$ are estimated using
a one dimensional Ising chain model \cite{Ising1925}.
The entropy per ion is approximated by
\[
S=-(3k_B/\mu_{\rm eff}^2)\,
\left[G(|\vec{M}_A|)+G(|\vec{M}_B|)\right]/2,
\]
with  
$G(M)= -\mu_{\rm eff}^2\,\ln\,2  - \MS^2\,\ln(1 - M^2/\MS^2)/2$
and the saturated magnetization $\MS= 7 \mu_B$,
as described in Appendix~\ref{appentropy}, Eqn.~(\ref{svonm}).
Gd$^{3+}$-ions exhibit a half filled 4f-shell
and a pure spin moment of $S=7/2$, such that  $L=0$ and  $J=7/2$.
The spin of the  Gd$^{3+}$-ion is described as an 8 niveau system. 
We simulate 256 Gd$^{3+}$-ions for one sublattice 
and count the number of possible states
for a particular magnetization $\vec{M}$.
The resulting entropy is approximated well by $G(|\vec{M}|)$ up to
0.7 $\MS$. For details see App.~\ref{appentropy}.

We assume an arbitrary analytic interaction energy
$\Phi(\vec{M}_A,\vec{M}_B)$, where 
$\vec{M}_A$ and $\vec{M}_B$ are two dimensional vectors 
restricting the mean field parameters
to the basal plane. From its Taylor expansion 
\[\Phi(\vec{M}_A,\vec{M}_B)=E_{\rm FM} + E_{\rm AFM}  + \FA+\cdots\]
we figure out three particular terms. For more details see App.~\ref{appexchange}.
The energy of the exchange interaction within one sublattice 
\[E_{\rm FM}
=-(3k_B/\mu_{\rm eff}^2)\, 
  (\Theta_W+\Theta_N) (\vec{M}_A^2+\vec{M}_B^2)/8 \]
and between two sublattices is
\[E_{\rm AFM}
=-(3k_B/\mu_{\rm eff}^2)\,
(\Theta_W-\Theta_N) (\vec{M}_A\cdot\vec{M}_B)/4.\]
Both polynomial terms are of lowest order in the mean-field parameters
restricted to the basal plane. They are isotropic due to 
the tetragonal symmetry.  The prefactors are choosen, such that
$\Theta_W$ is the Weiss temperature and 
$\Theta_N$ the \neel temperature. The anisotropic term
$\FA=f_a(\vec{M}_A)+f_a(\vec{M}_B)$
with
$f_a(\vec{M})=\CA\,M_x^2 M_y^2/4$ 
is of lowest order in $\vec{M}$, satisfying the tetragonal symmetry.
In the frame of our coordinate system the choice of $\CA>0$ is 
adapted to the direction of the magnetization.
The parameters $\Theta_{\rm N}, \Theta_{W}, \mu_{\rm eff}$ 
are known from the literature, whereas $\CA$ is a characteristic parameter
by which the textbook example is extended by our approach.

The AFM ordering parameter is defined by
\[\vec{D}:=(\vec{M}_A-\vec{M}_B )/2=D\,\vec{e}\platz\text{with}
\platz\vec{e}=(\cos\,\varphi,\sin\,\varphi),\]
where $\varphi$ is its angle with the $[110]$-direction.
The angle between the external field $\vec{B}$ 
and the $[110]$-direction is denoted by $\vartheta$, such that
$\vec{B} = B\,(\cos\,\vartheta\, ,\, \sin\,\vartheta ),$
as shown in Fig~{\ref{coord}}.

\subsection{Domains without external field}
First, we consider the case where no external field is applied.
The mean value of the magnetization per ion is $\vec{M}=0$,
and therefore,
$\vec{M}_A=-\vec{M}_B =\vec{D}$.
The free energy, Eqn. (\ref{masterfree}), becomes
\[ F(\vec{M}_A, \vec{M}_B)= F(\vec{D},-\vec{D})=F_0(D)+\FA(\vec{D})\]
with
\begin{eqnarray}
F_0(D):=( 3k_BT / \mu_{\rm eff}^2)
\left[G(D) - \Theta_N\,\vec{D}^2/(2T)
\right]
\end{eqnarray}
and $\FA=\,\CA\,D_x^2\,D_y^2/2$.
It is a function of the ordering parameter $\vec{D}$.
Above the \neel temperature $T\geq \Theta_N$, 
$F$ has a minimum for $\vec{D}=0$.
This is the paramagnetic regime, where the AFM ordering parameter vanishes.
For $T\leq \Theta_N$,
there are four  minima:
\[\vec{D}^I=(0,D_0),\;\;
\vec{D}^{II}=(D_0,0),\;\; 
-\vec{D}^I,\;\;-\vec{D}^{II}\]
with $D_0:= \MS \,\sqrt{1 -  T/\Theta_N}$ \label{dnull}. A detailed calculation can be found in App.~\ref{withoutfield}.
From the macroscopic point of view, the two mimima 
$\pm \vec{D}^I$ of domain~I with 
the magnetic moments in $[\bar{1}10]$- 
and  $[1\bar{1}0]$-direction are undistinguishable. 
This holds for domain~II in a similar way.

\subsection{Small field approximation}
Small external fields cause a small magnetization $\vec{M}$.
To describe our low field magnetization data the expansion 
of $F(\vec{M},\vec{D})$ up to second order in $\vec{M}$ is sufficient.
The calculation in App.~\ref{Fexpansion} yields
\[
F(\vec{M},\vec{D})
&=&F_0(D) + \vec{M}\,\hat{\chi}^{-1}\,\vec{M}/2
-\vec{M}\cdot\vec{B}+\FA,\,\text{with}
\\
\hat{\chi}^{-1}
:&=&
\chi_\perp^{-1} \,(1-\vec{e}\otimes \vec{e})
+\chi_\parallel^{-1}\,\vec{e}\otimes \vec{e},
\\
\hat{\chi}_\perp^{-1}
:&=&
(3k_B/\mu_{\rm eff}^2)\,\left[\frac{T\,G^\prime(D)}{D} - \Theta_W \right],
\\
\hat{\chi}_\parallel^{-1}
:&=&(3k_B/\mu_{\rm eff}^2)\left[T\,G^{\prime\prime}(D) - \Theta_W \right]
.\]
Since $M$ is small, we replace $D$ by $D_0=\MS\,\sqrt{1-T/\Theta_N}$, 
that is the value at the minimum, if no external field is applied.
With the second derivative
\[
G^{\prime\prime}(D_0)=\left(\Theta_N/T\right)^2
\left[2- T/\Theta_N\right]
\]
we get 
\begin{eqnarray}
\chi_\parallel 
&=&
\mu_{\rm eff}^ 2\;
[ 3 k_B\, \Theta_N\,(2\,\Theta_N-T )/T - 3 k_B\, \Theta_W ]^{-1},
\nonumber\\
\chi_\perp&=&\mu_{\rm eff}^ 2\;[ 3 k_B (\Theta_N-\Theta_W)\,]^{-1}
\label{xparallel}
\end{eqnarray}
such that the free energy depends only on the
magnetization and the direction of the ordering parameter:
$F=F(\vec{M}, \varphi)$.
Now we neglect $M$ in the anisotropic term, and get 
\[\FA(\varphi)
&=&
(\chi_\perp-\chi_\parallel)\,(B_{\rm sf}^ 2/8)\,\sin^22\varphi 
\label{phiA}\]
where we have introduced the spin-flop field
\[B_{\rm sf}^ 2:=\CA\,D_0^4 /(\chi_\perp-\chi_\parallel).\]

The next step is to eleminate $\vec{M}$.
A local minimum of $F(\vec{M},\varphi)$ is at
$\vec{M}=\hat{\chi}\,\vec{B}$.
By replacing $\vec{M}$ by its minimum
the susceptibility term 
$ \vec{M}\,\hat{\chi}^{-1}\,\vec{M}/2$
merges with the Zeeman term
and the free energy becomes
\[&&F(\varphi)
=F_0(D_0)-\vec{B}\,\hat{\chi}(\varphi)\,\vec{B}/2+  \FA(\varphi)
.\]
Now we have a closer look at the second term.
We introduce 
\[u:=-\cos(2\vartheta-2\varphi)\]
to parametrize the angle between the
ordering parameter $\vec{D}$ and  the external field
 $\vec{B} = B\,(\cos\,\vartheta\, ,\, \sin\,\vartheta )$.
We have $\vec{e}\cdot \vec{B}=B\,\cos(\vartheta-\varphi)$
and $(\vec{e}\cdot \vec{B})^2=B^2\,(1-u)/2$.
Since
\begin{eqnarray}
\vec{M}=
\hat{\chi}\,\vec{B}
&=&
\chi_\perp \,(1-\vec{e}\otimes \vec{e})\,\vec{B}
+\chi_\parallel\,\vec{e}\otimes \vec{e}\,\vec{B}
\end{eqnarray}
the susceptibility term becomes 
\[
\vec{M}\cdot\vec{B}=
\vec{B}\,\hat{\chi}(\varphi)\,\vec{B}
=B^2\,(\chi_\perp + \chi_\parallel)/2
+ B^2\,(\chi_\perp - \chi_\parallel)\,
u/2
.\]
The free energy becomes 
\begin{eqnarray}
F(\varphi)
=F_0(D_0)
&&
-B^2\,(\chi_\perp + \chi_\parallel)/4
- B^2\,(\chi_\perp - \chi_\parallel)\,u/4 \nonumber
\\&&
+(\chi_\perp-\chi_\parallel)\,(B_{\rm sf}^ 2/8)\,\sin^22\varphi.\end{eqnarray}
We introduce the reduced potential
\[\tilde{\phi}(\varphi)
:=-\left(B/B_{\rm sf}\right)^2\,u
+ (1/2) \,\sin^2(2\varphi),\]
which is the essential part of the free energy
\begin{eqnarray}
F(\varphi)
&=&F_0(D_0)-(B^2/4)\,(\chi_\perp+\chi_\parallel )\nonumber
\\&&
+ (B_{\rm sf}^2/4)\,(\chi_\perp-\chi_\parallel)\,\;\tilde{\phi}(\varphi).
\label{potphi}
\end{eqnarray}
The susceptibility term which originates from the AFM ordering is $\pi$-periodic
and the anisotropic term is $\pi/2$-periodic
with respect to $\varphi$.
So far, the problem is reduced to the task 
to find the minimum of a function that is the superposition
of a sine and a sine with the double frequency,
where the relative phase 
stems from the direction of the external magnetic field.
This problem has an algebraic solution only for selected angles.
In the reduced potential the relevant dynamics is visible.
The magnetic moments of the two sublattices
are fixed with respect to each other. The remaining degree of freedom is the rotation of the moments.

There are two competing forces:
The external magnetic field which
tries to turn the magnetic moments perpendicular to the field direction and
the anisotropy which tries to turn the moments along the $[110]$-direction.


\subsection{Field in $[100]$-direction}
For $\vartheta=\pi/4$ we have $u=-\sin\,2\varphi$ and
the reduced potential becomes
$\tilde{\phi}=-\left(B/B_{\rm sf}\right)^2\,u + u^2/2$.
It has a minimum at
$u_{\rm min}=(B/B_{\rm sf})^ 2 $ for $ B<B_{\rm sf}$ and
$u_{\rm min}=1$ otherwise. So we can fit the data on
\begin{eqnarray}
(\vec{M}\cdot\vec{B})/\vec{B}^ 2
=(\chi_\perp + \chi_\parallel)/2
+ (\chi_\perp - \chi_\parallel)\,
u _{\rm min}/2 
\label{Dir100Fit}
\end{eqnarray}
as shown in Fig.~\ref{fit100}.
At 3.5\,K we obtain 
$\chi_\perp=0.149\,\mu_B/$T,
$\chi_\parallel=0.0036\,\mu_B/$T and
$B_{\rm sf}=0.92\,$ T.
When increasing the external field,
the directions of the ordering parameters
turn perpendicular towards the external magnetic field direction, 
until  $B$  reaches $B_{\rm sf}$ where $M(B)$ shows a sharp kink.
Above  $B_{\rm sf}$, domain I and II 
are undistinguishable resulting in a field-independent susceptibility.

\subsection{Field in $[110]$-direction}
For $\vartheta=0$, we have $u=-\cos\,2\varphi$ and the  
reduced potential becomes
$\tilde{\phi}=-\left(B/ B_{\rm sf}\right)^2 \; u +(1-u^2)/2$.
It has an absolute minimum at
$u^I=1$, that we assign to domain~I.
In domain~I,
$\vec{M}_A$ and  $\vec{M}_B$ are perpendicular to the external field.
For $B<B_{\rm sf}$ there is a second, but local, minimum at
$u^{II}=-1$, that we assign to domain~II.
In the domains~I the moments rotate towards the
external fields, while in
the  domains~II the magnetizations of 
the sublattices $\vec{M}_A$ and  $\vec{M}_B$ 
keep their directions, but $\vec{M}_A$  becomes 
larger while $\vec{M}_B$ decreases. 
When increasing the external field a domain flip occurs.
This means that domain~II with relative volume size $a^{II}$, with the magnetic moments aligned 
parallel to $\vec{B}$, disappears and domain~I with relative volume size $a^{I}$, with the magnetic moments aligned 
perpendicular to $\vec{B}$, survives.

To predict the  relative volume sizes of the two domains I and II,
we use the infinite Ising chain \cite{Ising1925}. 
Every vertex of the Ising chain 
has two possible states with the energy difference 
$\Delta E$. If two neighbouring vertices have different states,
an additional wall energy contribution  $E_W$ is assumed, 
for details see App.~\ref{isinglabel}.
This yields the relative volume size of domain I to be
\begin{eqnarray}
a^I&=&\frac12 + \frac12\,\frac{\sinh\frac{ \beta \, \Delta E}{2}}
{\sqrt{\sinh^2 \frac{ \beta\,\Delta E }{2} + 
\frac{(\beta \EDF)^2}{4}  }}
\nonumber\\
&\simeq &
\frac12 + \frac12\,\Delta E/ \sqrt{\Delta E^2 + \EDF^2 }
\label{ising1}
\end{eqnarray}
with
$\EDF:=2\,k_BT\,\exp(-E_W/k_BT)$. From  Eqn.~(\ref{ising1}), with
$\Delta E=B^2(\chi_\perp-\chi_\parallel)/2$,
we get 
\begin{eqnarray}
(\vec{M}\cdot\vec{B})/ \vec{B}^2
&=&
\frac12(\chi_\perp + \chi_\parallel)
+ \frac12\,( \chi_\perp - \chi_\parallel)\,(\,u^I a^I + u^{II}a^{II}\,)\nonumber
\\
&=&
\frac12(\chi_\perp + \chi_\parallel) 
+ \frac12\,\frac{ \chi_\perp - \chi_\parallel}
{\sqrt{1+\left(\frac{B_{\rm df}}{B}\right)^4}}
\label{formula110}
\end{eqnarray}
for the two domains with a characteristic domain flip field
$ B_{\rm df}^2:=2\EDF / (\chi_\perp - \chi_\parallel)$.
Our data with the fit, using Eqn.~(\ref{formula110}), are shown in Fig.~\ref{22fitbild}.
From this we obtain
$\chi_\perp =0.149 \,\mu_B/$T,
$\chi_\parallel= 0.012\,\mu_B/$T,
$B_{\rm df}$=0.077\,T and $E_W$=3.1\,meV.

\subsection{Field in $\vartheta=22.5^\circ$-direction}

In contrast to the $[110]$ and $[100]$-direction, 
for an arbitrary in-plane direction spin-flop as well as domain flip occure.
For the external field 
22.5$^\circ$($\vartheta=\pi/8$) 
between the $[100]$- and $[110]$-direction,
the reduced potential becomes
$\tilde{\phi}=-\xi\, u  -\sqrt{u^2-u^4}/2 + 1/4$
with $\xi=\left(B/B_{\rm sf}\right)^2$.
The minima for domain I and II, calculated in App.~\ref{22grad}, are
\begin{eqnarray*}
u^I(\xi)&=&\sqrt{(1-\xi^2)/2+\sqrt{\xi^2\,\left(1+\xi^2/2\right)/2 }},
\nonumber\\
u^{II}(\xi)&=&-\sqrt{(1-\xi^2)/2-\sqrt{\xi^2\,\left(1+\xi^2/2\right)/2 }}.
\end{eqnarray*}
The magnetic energy per ion in domain~I is
\[E^I
=-\vec{M}^I\cdot \vec{B}/2
=-\left\{(\chi_\perp + \chi_\parallel)
+(\chi_\perp - \chi_\parallel)\,u^I\,  \right\}\;B^2/4\]
and in domain~II respectively. We feed Eqn.~(\ref{ising1}) with
\[
\Delta E= E^{II} - E^{I}
=(\chi_\perp-\chi_\parallel)\;(u^I-u^{II})\;B^2 /4
\]
to achieve 
\begin{eqnarray}
(\vec{M}\cdot\vec{B}) /\vec{B}^2 
&=&
\frac12 {(\chi_\perp+\chi_\parallel)}
+\frac12 {(\chi_\perp-\chi_\parallel)}\,(\,u^I\,a^I+u^{II}\,a^{II}\,)
\nonumber\\
&=&
\frac12 {(\chi_\perp+\chi_\parallel)}
+\frac14\,(\chi_\perp-\chi_\parallel)\,(u^I+u^{II})
\nonumber\\
&&+\frac14\,
\frac{(\chi_\perp-\chi_\parallel)\, ( u^I - u^{II} ) }
{\sqrt{1 + \left(\frac{B_{\rm df}}{B}\right)^4\,
           \left(\frac{2}{u^I - u^{II}}\right)^2
      }
}
\label{22fit}
\end{eqnarray}
for fitting.
Inserting $B_{\rm df}$, $B_{\rm sf}$, and the average values 
of $\chi_\perp$, $\chi_\parallel$ determined for the measurements
on the $[100]$- and the $[110]$-direction, Eqn.~(\ref{22fit}) fits the 
measured data at $T=3.5\,\rm K$ with $\vec{B}$ parallel to the 
22.5$^{\circ}$-direction Fig.~\ref{22fitbild}.

\begin{figure}
\centering
\includegraphics[width=0.5\textwidth]{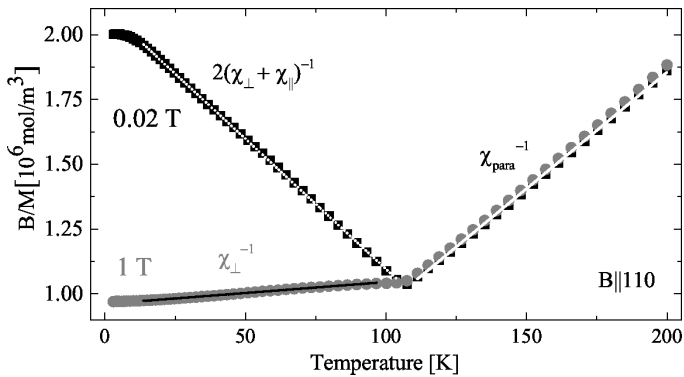}
\caption[]{
For $\vec{B}\parallel 110$, the temperature dependence of $B/M$ for $0.02\,$T 
is shown with black squares and for a higher field with grey circles.
Both measured curves coincide above the \neel temperature.
The figure shows three lines meeting at the magnetic phase transition,
that can be fitted by three different formulae.
For low fields, between  15\,K and $\Theta_N$, $B/M$ can be predicted
by Eqn.~(\ref{chiinverse}), that is plotted by a white dotted line.
For higher fields (black solid line), we utilize Eqn.~(\ref{chpe}).
Above the \neel temperature, in the paramagnetic regime,
$B/M$ is field independent (white solid line)
and is described by Eqn.~(\ref{chpa}).
}
\label{rezchi}
\end{figure}

\subsection{Temperature dependence of the susceptibility}
For very low external fields applied 
in an in-plane direction shown in Fig.~\ref{fit110bild},
the macroscopic susceptibility
$M/B$ averaged over many domains is isotropic.
This is a consequence of the macroscopic fourfold symmetry,
and the assumption that $M(B)$ is an analytic function in the rigorous sence \cite{RAF}.
For higher fields, when $\vec{D}$ is perpendicular to
$\vec{B}$ the ratio $M/B$ 
becomes isotropic and constant (Fig.~\ref{fit100}).
The in-plane anisotropic effects occur in the interime.
Neglecting $\Theta_W=(8\pm 5)\,$K \cite{Kliemt2015},
Eqn.~(\ref{xparallel}) implies
\begin{eqnarray}(\chi_\perp+\chi_\parallel)/2
=\mu_{\rm eff}^ 2 [ \,  3 k_B(2\Theta_N-T) \,]^{-1},
\label{chiinverse}
\end{eqnarray}
for external fields below 0.02 T and
\begin{eqnarray}
\chi_\perp=\mu_{\rm eff}^ 2/( 3 k_B\,\Theta_N)
\label{chpe}
\end{eqnarray}
for intermediate fields. 
The paramagnetic behaviour  above $\Theta_N$ is described by
\begin{eqnarray}
\chi_{\rm para}=\mu_{\rm eff}^ 2/ ( 3 k_B \,T).
\label{chpa}
\end{eqnarray}
The three lines in Fig.~\ref{rezchi} meet at $\Theta_N$.

\section{Summary}
GdRh$_2$Si$_2$ single crystals have been investigated 
by VSM, XMCD and X-ray resonant magnetic scattering. 
In the course of our low temperature investigation, we found 
that this compound represents an exemplary case 
for a simple antiferromagnetic order with in-plane ordered moments
and weak in-plane anisotropy. 
Applying the field parallel to the tetragonal plane,
$M/B$ is isotropic in the limit $B=0$. 
When increasing the field, strong anisotropic effects arise.
For field parallel to the $[100]$-direction,
we observed a spin-flop transition which is 
the rotation of local spin directions \cite{Stryjewski1977}.
In contrast, we explain the  sudden increase of the susceptibility in $[110]$-direction
at low fields by domain effects. 
To our knowledge, such a direct measurement of the change 
in the domain distribution in a tetragonal compound has 
not been reported, so far. 
Since the low field regime yields a magnetic behaviour 
of a unique simplicity, 
we set up a magnetic mean field model combined 
with an Ising chain model.
With the assumption that the magnetic moments are aligned
parallel to the $[110]$-direction,
the experimental data were  perfectly reproduced.
We therefore conclude that the magnetic moments are aligned 
along the $[110]$-direction in the tetragonal 
lattice at low temperatures.
Additional magnetic scattering experiments confirm 
the arrangement of the magnetic moments 
in staggered ferromagnetic layers along the $[001]$-direction.
The magnetization of these layers have opposite directions
forming the antiferromagnetic bulk.
The XMCD study rules out a Rh contribution to the magnetization,
proposed in Ref.~[\onlinecite{Felner1984}]. 
The slight enhancement of the effective moment observed in the susceptibility data is likely due to the polarization of the 5d-Gd electrons due to on-site Hunds coupling to the 4f electrons.

\begin{acknowledgments}
 KKl and CK grateful acknowledge support by the DFG through grant SFB/TR49.
\end{acknowledgments}

\appendix
\setcounter{equation}{0}


\section{Magnetic scattering\label{app:magnetic_scattering}}

The magnetic scattering intensity for the employed scattering geometry and incident $\pi$ polarisation can be written, up to some scale factor, as \cite{hill1996-ac}
\begin{equation}
\begin{split}
I_{\pi} & = I_{\pi\sigma'} + I_{\pi\pi'} \\ 
& = \left|-M_{(110)}\cos\theta+M_{(00\bar{1})}\sin\theta\right|^2 + \left|M_{(1\bar{1}0)}\sin 2\theta\right|^2.\nonumber
\end{split}
\end{equation}
With $M_{(00\bar{1})} = 0$ and $\sin 2\theta=2\sin\theta\cos\theta$ we obtain
\begin{equation}
I_{\pi} = \left(M_{(110)}^2 + 4M_{(1\bar{1}0)}^2\sin^2\theta\right)\cos^2\theta .\nonumber
\end{equation}
For GdRh$_2$Si$_2$ the $c$ lattice constant is 9.986\,\AA\ and the photon energy of the Gd $M_5$ absorption edge is $E = 1183\,$eV corresponding to an X-ray wavelength of $\lambda=10.48\,$\AA. This yields for the (0\,0\,1) reflection $\sin\theta = \lambda/(2c) \approx 1/2$ which if put into the above equation results in
\begin{equation}
I_{\pi} = \left(M_{(110)}^2 + M_{(1\bar{1}0)}^2\right)\cos^2\theta = M_{ab}^2\cos^2\theta\nonumber
\end{equation}
with $M_{ab}$ the absolute value of the moment in the basal plane. Under the conditions used in this experiment and for $M_{(001)} \ll M_{ab}$ the scattering intensity is therefore independent of the orientation of the moment in the $ab$ plane and thus also independent of any domain structure. It is only proportional to the staggered magnetization.

\section{Entropy}\label{appentropy}

\begin{figure}
\centering
\includegraphics[width=0.5\textwidth]{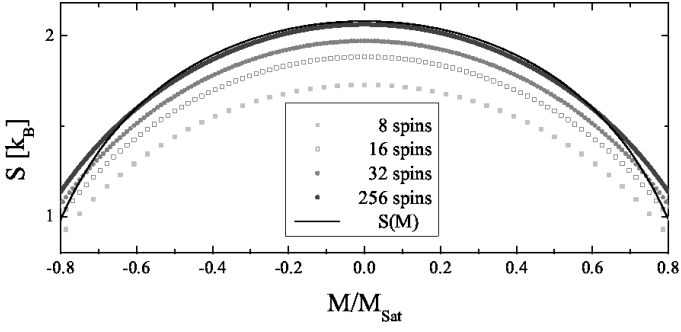}
\caption[]{
The numerical simulation of the entropy for a fixed numbers of spins
(grey squares) is approximated by $S(M)$ Eqn.~(\ref{svonm}).
}
\label{approxsim}
\end{figure}

To determine the entropy of a single ion,
we perform a numerical simulation.
We fix the number of spins to be $n=8$, $16$, $64$ and $256$.
Every spin has 8 possible states, with magnetic moments 
\[m_i = -7\,\mu_B,-5\,\mu_B,\cdots,7\,\mu_B.\]
The number of all states of this ensemble is $8^n$.
Every ion $i=1,\cdots,n$ has a sharp magnetic moment $m_i$.
For every state we determine the mean magnetization per ion by
\[M=\frac{1}{n}\sum_{i=1}^n  m_i.\]
We count the number $N(M)$ of states for every possible $M$.
For $M=\MS=7\,\mu_B$ there is only one state.
The entropy per moment is \[S(M)=k_B\,\ln\,N(M).\] 
For $M=0$ the entropy is maximal for any even $n$.
Consider an system with an infinite number of spins.
If every state has the same probability, then $M=0$.
In particular, every ion has 8 states with the same probability.
In this case, the entropy per ion
takes the value $S(M)|_{M=0}=k_B\,\ln \,8$.
With $\MS=7\mu_B$ and $\mu_{\rm eff}\simeq 8.28\,\mu_B$,
the entropy  
\begin{equation}
S(M)=k_B\,\left( \ln\,8 + \frac32\,\frac{\MS^2}{\mu_{\rm eff}^2}\,\ln
\left(1-\frac{M^2}{\MS^2}\right) \;\right) 
\label{svonm}
\end{equation}
used in our model,
fits the simulation well, which is demonstrated in Fig.~\ref{approxsim}.

\section{Interaction terms}
\label{appexchange}
The interactions between and within the sublattices are described by an 
analytic function as a contribution to the free energy Eqn.~(\ref{masterfree}).
We therefore expand an arbitrary analytic function in its Taylor series
\[&&\phi(M_{Ax},M_{Ay}, M_{Bx},M_{By})\\
&=&\sum_i \sum_j \sum_k\sum_l
c_{ijkl} M_{Ax}^i M_{Ay}^j M_{Bx}^k M_{By}^l.\]
For the Taylor coefficients $c_{ijkl}$,
the following symmetry conditions are required
in a tetragonal compound:
\begin{enumerate}
\item  Changing the sublattices requires  $c_{ijkl}=c_{klij}$.
\item  The x-axis can be changed with the y-axis, so $c_{ijkl}=c_{jilk}$.
\item  The reflection symmetry on the $y$-axis yields 
       that $i+k$ has to be even.
\item  The reflection symmetry on the $x$-axis yields,
       that $j+l$ has to be even.
\end{enumerate}
From 3 and 4, we deduce that the order $i+j+k+l$ is even,
otherwise $c_{ijkl}$ will be zero.
These four symmetry conditions yield for the second order terms that
\[ c_{1010}&=&c_{0101}, \\
c_{2000}&=&c_{0200}=c_{0020}=c_{0002}, \]
and for the fourth order terms that
\[c_{2002}&=&c_{0220},\\
c_{2020}&=&c_{0202},\\
c_{2200}&=&c_{0022},\\
c_{3010}&=&c_{1030}=c_{0103}=c_{0301},\\
c_{4000}&=&c_{0400}=c_{0040}=c_{0004}.\]
The Taylor series becomes
\[\phi(M_{Ax},M_{Ay}, M_{Bx},M_{By})
=c_{0000} + \phi^{(2)} + \phi^{(4)} + \cdots  \]
with the second order terms
\[\phi^{(2)}&=&c_{1010} (M_{Ax}\cdot M_{Bx} + M_{Ay}\cdot M_{By})\\
&&+c_{2000} ( M_{Ax}^2 + M_{Ay}^2 + M_{Bx}^2 + M_{By}^2)\]
and fourth order terms
\[\phi^{(4)}&=&
c_{1111} (M_{Ax}\cdot M_{Ay} \cdot M_{Bx}\cdot M_{By})
\\&&
+c_{2002} ( M_{Ax}^2 \cdot M_{By}^2 + M_{Ay}^2 \cdot M_{Bx}^2)
\\&&
+c_{2020} ( M_{Ax}^2 \cdot M_{Bx}^2 + M_{Ay}^2 \cdot M_{By}^2)
\\&&
+c_{2200} ( M_{Ax}^2 \cdot M_{Ay}^2 + M_{Bx}^2 \cdot M_{By}^2)
\\&&
+c_{3010} ( M_{Ax}^3 \cdot M_{Bx} + M_{Ax}\cdot M_{Bx}^3 
\\&&\hspace{0.5cm}           + M_{Ay}^3 \cdot M_{By} + M_{Ay}\cdot M_{By}^3 )
\\&&
+c_{4000} ( M_{Ax}^4 + M_{Ay}^4 + M_{Bx}^4 + M_{By}^4)
\]
with these symmetries. The constant term has no physical effect.

\subsubsection{Second order terms}
There remain two second order terms
\[E_{\rm AFM}&=&  c_{1010} (M_{Ax}\cdot M_{Bx} + M_{Ay}\cdot M_{By})\\
&=& c_{1010}\,\vec{M}_A\cdot\vec{M}_B\]
and
\[E_{\rm FM}&=&c_{2000} ( M_{Ax}^2 + M_{Ay}^2 + M_{Bx}^2 + M_{By}^2)\\
&=&c_{2000}\,(\,\vec{M}_A^2+\vec{M}_B^2\,).\]
Since both are isotropic, we need terms of higher order for the in-plane anisotropy. 
Now, we will show that
\[c_2:= c_{2000}=-(3k_B/\mu_{\rm eff}^2)\,(\Theta_W+\Theta_N)/8\]
and
\[c_1:= c_{1010}=-(3k_B/\mu_{\rm eff}^2)\,(\Theta_W-\Theta_N) /4.\]
If neglecting the anisotropic term, the free energy becomes
\[&&F(\vec{M}_A,\vec{M}_B)\\
&=&-T\;S + E_{\rm FM} + E_{\rm AFM} - \vec{M} \cdot\vec{B} \\
&=&- T\;S +c_2\,(\,\vec{M}_A^2+\vec{M}_B^2\,)
   + c_1\,\vec{M}_A\cdot\vec{M}_B \\
&& - ( \vec{M}_A+ \vec{M}_B   )  \cdot\vec{B}/2.\]
The entropy of a single ion is approximated by
\[S=-(3k_B/\mu_{\rm eff}^2)\,
\left[G(|\vec{M}_A|)+G(|\vec{M}_B|)\right]/2,\]
with
\[G(M)= -\mu_{\rm eff}^2\,\ln\,2  - \MS^2\,\ln(1 - M^2/\MS^2)/2.\]
The mimimum conditions are
\[\frac{\partial F}{\partial \vec{M}_A} 
&=&
(3k_B\,T/\mu_{\rm eff}^2)\,
\frac{\vec{M}_A /2  }{1 -  M_A^2/\MS^2}\\
&& \hspace {0.5cm}+ 2 c_2 \,\vec{M}_A+ c_1 \,\vec{M}_B-\vec{B}/2=0,\\
\frac{\partial F}{\partial \vec{M}_B}
&=&
(3k_B\,T/\mu_{\rm eff}^2)\,
\frac{\vec{M}_B /2  }{1 -  M_B^2/\MS^2}\\
&& \hspace {0.5cm}
+ 2 c_2 \,\vec{M}_B+ c_1 \,\vec{M}_A-\vec{B}/2=0.\]
In the paramagnetic regime, where $\vec{M}=\vec{M}_A=\vec{M}_B$,
both minimum conditions are equivalent, and yield
\[(3k_B\,T/\mu_{\rm eff}^2)\,\frac{\vec{M}   }{1 -  M^2/\MS^2}
+ ( 4 c_2 + 2 c_1) \,\vec{M}=\vec{B}.\]
For small external fields, we have small magnetization such that
\[\vec{M}
&=&\frac{1}{ (3k_B\,T/\mu_{\rm eff}^2) + ( 4 c_2 + 2 c_1)}\,\vec{B}\\
&=&\frac{\mu_{\rm eff}^2}{3k_B}\,\frac{1}{T-\Theta_W}\,\vec{B},\]
hence
\[(4c_2+2 c_1)=-(3k_B\,\Theta_W/\mu_{\rm eff}^2).\]
In the antiferromagnetic regime there is no external field and therefore
$\vec{D}=\vec{M}_A =- \vec{M}_B $. Both minimum conditions become
\[ (3k_B\,T/\mu_{\rm eff}^2)\,
\frac{\vec{D}}{1-D^2/\MS^2}+(4c_2 - 2c_1)\,\vec{D}=0.\]
There is the solution  $\vec{D}=0$ for the paramagnetic case.
With the assumption $\vec{D}\neq 0$, the condition reduces to
\[(3k_B\,T/\mu_{\rm eff}^2)\,+(4c_2-2c_1)\,(1-D^2/\MS^2)=0\]
and then to
\[\frac{(3k_B\,T/\mu_{\rm eff}^2)}{4c_2-2c_1}+1=D^2/\MS^2.\] 
In the limit $\vec{D}=0$, we approach the 
antiferromagnetic phase transition. This yields
\[\frac{(3k_B\,\Theta_N/\mu_{\rm eff}^2)}{4c_2-2c_1}+1=0\]
for the determination of the \neel temperature. Now we have two equations
\[(4c_2+2c_1)&=&-(3k_B\,\Theta_W/\mu_{\rm eff}^2),\\
  (4c_2-2c_1)&=&-(3k_B\,\Theta_N/\mu_{\rm eff}^2),\]
to determine $c_1$ and $c_2$. The sum of both yields
\[c_2=-(3k_B/\mu_{\rm eff}^2)\,(\Theta_W+\Theta_N)/8\]
and the difference
\[c_1=-(3k_B/\mu_{\rm eff}^2)\,(\Theta_W-\Theta_N)/4.\]

\subsubsection{Forth order terms}
We consider small $\vec{M}$, and therefore neglect it in terms of 
fourth order. This means that $M_{Ax}=-M_{Bx}=D_x$ and
 $M_{Ay}=-M_{By}=D_y$. The fourth order terms become
\[\phi^{(4)}&=&
(\,c_{1111}+2\,c_{2200}\,+2\,c_{2002}\,)(\,D_{x}^2\cdot D_{y}^2)\\
&&+(\,c_{2020}+2\,c_{3010}\,+2\,c_{4000})(\,D_{x}^4+D_{y}^4).\]
This can be re-written as
\[\phi^{(4)}&=&
(\,c_{1111}+2\,c_{2200}\,-\frac12 c_{2020}+c_{3010}\,+c_{4000}) 
(\,D_{x}^2\cdot D_{y}^2)\\
&&+(\,c_{2020}+2\,c_{3010}\,+2\,c_{4000})(\,D_{x}^2+D_{y}^2)^2.\]
The second summand is isotropic and of no further interest. 
So we use only the first term which can be rewritten as 
\[\FA&=&f_a(\vec{M}_A)+f_a(\vec{M}_B)\\
&=&\frac{\CA}{4}\,\left( \,M_{Ax}^2 M_{Ay}^2+M_{Bx}^2 M_{By}^2\right).\]


\section{Volume size of the domains modelled by an Ising chain}
\label{isinglabel}
To predict the  relative volume sizes of the two domains I and II,
we use the model of an Ising chain with $n$ links.
Every vertex of the Ising chain has two possible states 
with the energy difference 
$\Delta E$. If two neighbouring vertices have different states,
an additional wall energy contribution $E_W$ is assumed.
For a particular state $s$, the total energy has the form
\[E_s=-n\,(a_s^I-a_s^{II})\,\frac12\,\Delta E + E_s^{\rm wall},\]
where $E_s^{\rm wall}$ is the energy contribution of all walls and
$a_s^I$, $a_s^{II}$ are the relative volume sizes of domains I, II, 
respectively. The canonical partition function is
\[Z&=&\sum\limits_{s}\,e^{-\beta E_s}\\
&=&\sum\limits_{s}\,e^{n\,(a_s^I-a_s^{II})\,\alpha-\beta\,E_s^{\rm wall}}\]
with $\alpha=\beta\,\Delta E/2$. Thus, the expectation value 
for the difference of the relative volume sizes 
of domains I and domain II with respect to the canonical ensemble is
\[a^I-a^{II}&=&\frac{1}{Z}\sum\limits_{s}(a_s^I-a_s^{II})\,e^{-\beta E_s}\\
&=&\frac{1}{n}\,\frac{d}{d \alpha}\,\log\,Z.\]
In 1925, E. Ising published
an explicite formula for the canonical partition function
\[Z=\left(\cosh\,\alpha+\sqrt{\sinh^2\,\alpha
+e^{-2\beta\,E_W}}\right)^n+\cdots\]
for a chain with $n$ links \cite{Ising1925}.
This yields the difference of the relative volume 
between domain I and II to be
\[a^I-a^{II}&=&\frac{1}{n}\,\frac{d}{d \alpha}\,\log\,Z
= \frac{\sinh\, \alpha+\frac{ \sinh\,\alpha\;\cosh\,\alpha} 
  {\sqrt{\sinh^2\,\alpha + e^{-2\beta\,E_W}}}}
  {\cosh\,\alpha+\sqrt{\sinh^2\,\alpha+e^{-2\beta\,E_W}}}\\
&=&\frac{\sinh\,\alpha}{\sqrt{\sinh^2 \,\alpha + e^{-2\beta\,E_W}}}.\]
Since $a^I+a^{II}=1$ we have
\[a^I&=&\frac12 + \frac12\,(a^I-a^{II})\\
&=&\frac12+\frac12\,\frac{\sinh\frac{\beta\,\Delta E}{2}}
{\sqrt{\sinh^2\frac{\beta\,\Delta E}{2}+\frac{(\beta \EDF)^2}{4}}}\]
with $\EDF:=2\, k_BT \,\exp(-E_W/ k_BT)$. In this work, we use the approximation
\begin{eqnarray}
a^I&\simeq &
\frac12 + \frac12\,\Delta E/ \sqrt{\Delta E^2 + \EDF^2 }
\label{ising}
\end{eqnarray}
for small $\Delta E$. 


\section{Free energy expansion up to second order in \vec{M}}
\label{Fexpansion}

In this appendix, we will expand the free energy 
\[F(\vec{M}_A,\vec{M}_B)&=&
-T\;S +  E_{\rm FM} +  E_{\rm AFM} - \vec{M} \cdot\vec{B} + \FA\]
up to second order in the magnetization $\vec{M}$.
We have $\vec{M}_A = \vec{M} + \vec{D}$ and $\vec{M}_B = \vec{M} - \vec{D}$,
with 
\[\vec{D}=D\,\vec{e}=D\,(\cos\,\varphi\,,\,\sin\,\varphi).\]

\subsubsection*{Entropy}
We start with the Taylor expansion
\[
\sqrt{1+x}=1 + \frac12\,x - \frac{1}{8}\,x^2
\]
and get
\[&&
|\vec{D}+\vec{M}|
=\sqrt{D^2 + M ^2 +2\, \vec{D}\cdot\vec{M}   }
\\
&=&D\,\sqrt{ 1 + \frac {2}{D}\,\left( \vec{e}\cdot\vec{M} + \frac{1}{2D} M^2
\right)     }
\\
&\simeq&
D+\left( \vec{e}\cdot\vec{M} + \frac{1}{2D} M^2\right)     
-\frac1{2D}\,\left( \vec{e}\cdot\vec{M} + \frac{1}{2D} M^2\right)^2
\\
&\simeq&
D + \vec{e}\cdot\vec{M} +
\frac{1}{2D}\,\left[ M^2- (\vec{e}\cdot\vec{M})^2   \right]
\]
for the absolute function.
This gives us 
\[
 |\vec{D}+\vec{M}|   +  |\vec{D}-\vec{M}| - 2 D   
= \frac{1}{D}\,\left[  M^2-(\vec{e}\cdot\vec{M} ) ^2  
\right]+\cdots
\]
and
\[
 \left[|\vec{D}+\vec{M}| -D \right]  ^2
+
\left[ |\vec{D}-\vec{M}| -D \right]  ^2
= 2\,(\vec{e}\cdot\vec{M})^2 +\cdots
.\]
With the Taylor expansion
\[G(|\vec{D}+\vec{M}|)&=&
G(D) 
+  G^\prime(D) (|\vec{D}+\vec{M}| -D   )
\\ && \vspace{1cm}
+  G^{ \prime\prime} (D)\,\frac12\, \left[|\vec{D}+\vec{M}| -D \right]  ^2+\cdots   
\]
we get
\[
&&G(|\vec{D}+\vec{M}|)+
G(|\vec{D}-\vec{M}|)
\\
&=&
2\,G(D) 
+  G^\prime(D) 
 \frac{1}{D}\,\left[ M^2-(\vec{e}\cdot\vec{M} ) ^2 \right]
\\&&
+  G^{ \prime\prime} (D)\,(\vec{e}\cdot\vec{M})^2 +\cdots.
\]
Hence, the entropy term becomes
\[
S&=&-\KBMUEF\,\frac12\,
\left[G(|\vec{M}_A|)+G(|\vec{M}_B|)\right]
\\
&=&-\KBMUEF\,\frac12\,
\left[G(|\vec{M}+\vec{D}|)+G(|\vec{M}-\vec{D}|)\right]
\\
&=&-\KBMUEF\,G(D) -
\KBMUEF\,
\frac12\,
\frac{  G^\prime(D)}{D}\,\left[M^2-
(\vec{e}\cdot\vec{M} ) ^2   \right]
\\
&& \hspace{1cm}
-\KBMUEF\,
\frac12\,
 G^{ \prime\prime} (D)\,(\vec{e}\cdot\vec{M})^2.
\]

\subsubsection*{Dipole Energy}
The energy of the dipole interaction within one sublattice
\[
E_{\rm FM}&=&-
\KBMUEF
\,\frac18\, 
  (\Theta_W+\Theta_N) (\vec{M}_A^2+\vec{M}_B^2)
\\
&=&-
\KBMUEF\,\frac14\,
  (\Theta_W+\Theta_N) (\vec{D}^2+\vec{M}^2)
\]
and between two sublattices is
\[E_{\rm AFM}
&=&-
\KBMUEF\,\frac14\,
(\Theta_W-\Theta_N) (\vec{M}_A\cdot\vec{M}_B)
\\
&=&-
\KBMUEF\,\frac14\,
(\Theta_W-\Theta_N) (\vec{M}^2-\vec{D}^2  )
,\]
and the sum is
\[
E_{\rm FM} + E_{\rm AFM}
&=&
-\KBMUEF\,\frac12\,
\left[
\Theta_W \vec{M}^2 + \Theta_N \vec{D}^2 
\right].
\]

\subsubsection*{Entropy and dipole interaction}
The sum of the entropy term and both dipole interaction terms becomes
\[
&&-T\cdot  S + E_{\rm FM} + E_{\rm AFM}
\\
&=& \frac{3k_BT}{\mu_{\rm eff}^2} \,G(D) +
\frac{3k_BT}{\mu_{\rm eff}^2}
\,\frac12\,
\frac{  G^\prime(D)}{D}\,\left[M^2-(\vec{e}\cdot\vec{M} ) ^2   
\right]
\\ && 
+
\frac{3k_BT}{\mu_{\rm eff}^2}
\,\frac12\,  G^{ \prime\prime} (D)\,(\vec{e}\cdot\vec{M})^2
\\
&&-
\KBMUEF\,\frac12\,
\left[
\Theta_W \vec{M}^2 + \Theta_N \vec{D}^2
\right]
\\
&=&
F_0(D) +
\frac12\,  
\vec{M}    
\,
\hat{\chi}^{-1}
\,
\vec{M}
\]
with
\begin{eqnarray}
F_0(D):=\frac{3 k_B T}{ \mu_{\rm eff}^2 }\,
\left[G(D) - \frac12 \, \frac{\Theta_N}{T}\, \vec{D}^2
\right],
\end{eqnarray}
\[  
\hat{\chi}^{-1}
&=&
\chi_\perp^{-1} \,(1-\vec{e}\otimes \vec{e})
+
\chi_\parallel^{-1}
\,\vec{e}\otimes \vec{e},
\]
with the dyadic product
\[
\vec{e}\otimes \vec{e}=
\begin{pmatrix}
\cos^2\varphi& \cos\,\varphi\;\sin\,\varphi \\
\sin\,\varphi\;\cos\,\varphi  & \sin^2\varphi
\end{pmatrix}
\]
and
\[
\hat{\chi}_\perp^{-1}
&=&
\KBMUEF \,
\left[T\,\frac{G^\prime(D)}{D} - \Theta_W \right],
\]
\[
\hat{\chi}_\parallel^{-1}
&=&\KBMUEF\,
\left[
T\,G^{\prime\prime}(D) - \Theta_W
\right]
.\]

\subsubsection*{Anisotropic contribution}
For the anisotropic term, we neglect any non zero order of $\vec{M}$.
So we get
\[\FA&=&\frac12\,\CA\,D_x^2 D_y^2
=\frac12\,\CA\,D^4 \,\cos^2\varphi\,\sin^2\varphi
\\
&=&\CA\,D^4\,\frac18\,\sin^22\varphi.\]

\subsubsection*{Free energy up to second order in $\vec{M}$}
We summarize the  expansion of  $F(\vec{M},\vec{D})$ 
up to second order in $\vec{M}$ by
\[
F(\vec{M},\vec{D})
&=&F_0(D) + \vec{M}\,\hat{\chi}^{-1}\,\vec{M}/2
-\vec{M}\cdot\vec{B}+\FA.\]


\section{Domains without external field}
\label{withoutfield}
We consider the case without any external field.
The magnetization $\vec{M}$ is zero.
The AFM ordering parameter is 
\[ \vec{D}:=\frac12\,(\vec{M}_A-\vec{M}_B )=D\,\vec{e}
\platz\text{with}\platz\vec{e}=(\cos\,\varphi,\sin\,\varphi). \]
For $\vec{M}=0$ the free energy reduces to
$F=F_0+\FA$ with
\begin{eqnarray}
F_0(D):=\frac{3 k_B T}{ \mu_{\rm eff}^2 }\,
\left[G(D) - \frac12 \, \frac{\Theta_N}{T}\, \vec{D}^2
\right]
\end{eqnarray}
and 
\[\FA=\frac12\,\CA\,D_x^2\,D_y^2\]
with $\CA>0$.
Now we want to find the minimum of the free energy with respect to
$\vec{D}$.
The absolute value of $\vec{D}$ at the minimum is
determined by the isotropic part alone,
since $\FA$ becomes zero for a particular directions. 
The extremum condition for the isotropic part is
\[
\frac{\partial F_0}{\partial D}
=
\frac{3 k_B T}{ \mu_{\rm eff}^2 }\,
\left[
G^\prime(D)  -  \frac{\Theta_N}{T}\,D
\right]=0
.\]
The derivative of
\[
G(D)= -\mu_{\rm eff}^2\,\ln\,2
- \MS^2\,\frac12\,\ln\left(1 - \frac{D^2}{\MS^2}\right)
\]
is
\[
G^\prime(D)
= \frac{D}{   1- \frac{D^2}{\MS^2} },
\]
and we get 
\[
\frac{\partial F_0}{\partial D}
=
\frac{3 k_B T}{ \mu_{\rm eff}^2 }\,
\left[
 \frac{D}{1- \frac{D^2}{\MS^2} } 
 -  \frac{\Theta_N}{T}\,D
\right].
\]
The extremum condition
$\partial F_0/ \partial D=0 $ yields
\[D=0\]
or
\[1 -\frac{ D^2}{\MS^2}=  \frac{T}{\Theta_N}.\]
Above the \neel temperature $T\geq \Theta_N$,
$D=0$  is the only physical solution.
The AFM ordering
parameter $\vec{D}$ remains zero.
For $T\leq \Theta_N$, 
$D=0$ becomes a maximum and the second possibility of the 
extremum condition is solvable:
\[
D=\MS\,\sqrt{1 - \frac{T}{\Theta_N}    }
.\]
The direction of $\vec{D}$ at the minimum is determined by the anisotropic term
\[\FA&=&\frac12\,\CA\,D_x^2 D_y^2
\]
with $\CA>0$.
There are four  minima:
\[\vec{D}^I=(0,D_0),\;\;
\vec{D}^{II}=(D_0,0),\;\; 
-\vec{D}^I,\;\;-\vec{D}^{II}\]
with 
\[D_0:= \MS \,\sqrt{1 -  \frac {T}{\Theta_N} }.
 \label{dnull}
\]


\section{Field in $\vartheta=22.5^\circ$-direction}
\label{22grad}

For $\vartheta=\pi/8$ we have
\[
u=-\cos(2\vartheta-2\varphi) 
 = -\sin \,\theta,\]
with $\theta:=2\varphi + \pi/4$. Moreover, we have 
\[\sin^2(2\varphi)
=\frac12 - \frac12\,\cos\left(2\theta - \frac{\pi}{2}\right)
=\frac12 -\frac12\,\sin(2\theta ),\]
and therefore the reduced potential becomes  
\[\tilde{\phi}
&=& - \xi\,u + \frac12 \,\sin^2(2\varphi)
\\
&=&  \xi\,\sin\,\theta - \frac14\,\sin\,2\theta +\frac14 \]
with $\xi:=(B/ B_{\rm sf} )^2$.
The extremum condition is
\[
\frac{ \partial \tilde{\phi}}{\partial\theta}
=\xi\,\cos\,\theta - \frac12\,\cos\,2\theta=0
\]
and becomes
\[
 0 =\cos^2\theta-\xi\,\cos\,\theta   -\frac12.
\]
The solutions for the extremum condition, if they exist, are
\[
 \cos\,\theta _{\rm ex}
&=& \frac{\xi}{2} \pm\sqrt {\frac{\xi^2}{4}  +\frac12  }
\\
&=&
\frac12 \left[ \xi\pm \sqrt{\xi^2+2}  \right].
\]
Their squares are
\[
\cos^2\theta_{\rm ex} 
=
\frac12(\xi^2+1) \pm \frac12\,\xi \, \sqrt{\xi^2+2}
\]
and hence
\[u^2_{\rm ex}=
 \sin^2\theta_{\rm ex}=
1 -  \cos^2\theta_{\rm ex} 
=
\frac12(1- \xi^2) \mp \frac12\,\xi \, \sqrt{\xi^2+2}.
\]
Next, we have to consider four extrema
$\theta_1,\cdots,\theta_4$.
The first and second extrema $\theta_1$ and $\theta_2$ are expressed by
\[
\cos\,\theta _1=\cos\,\theta_2
&=& \frac{\xi}{2}  - \sqrt {\frac{\xi^2}{4}  +\frac12  },
\]
\[
\sin\,\theta_1
&=&
\sqrt{ \frac12(1- \xi^2) + \frac12\,\xi \, \sqrt{\xi^2+2} },
\]
\[
\sin\,\theta_2         
&=&-
\sqrt{ \frac12(1- \xi^2) + \frac12\,\xi \, \sqrt{\xi^2+2} } 
\] 
and they exists for all $\xi\geq 0$.
The remaining extrema are expressed by
\[
\cos\,\theta _3=\cos\,\theta_4 
&=& \frac{\xi}{2} + \sqrt {\frac{\xi^2}{4}  +\frac12  },
\]
\[
\sin\,\theta_3
&=&
\sqrt{ \frac12(1- \xi^2) - \frac12\,\xi \, \sqrt{\xi^2+2} },
\]
\[
\sin\,\theta_4
&=&-
\sqrt{ \frac12(1- \xi^2) - \frac12\,\xi \, \sqrt{\xi^2+2} }.
\]
Since the argument in the root of the sinus can become negative,
$\theta_3$ and $\theta_4$ exists only for $\xi\leq\frac12$.
The minimum condition is
\[
\frac{ \partial^2 \tilde{\phi}}{\partial^2\theta }
=-\xi\,\sin\,\theta  + \sin\,2\theta
=
\sin\,\theta \, (-\xi+2\,\cos\,\theta)
>0.
\]
We have
$-\xi+2\,\cos\,\theta_{1,2} < 0$, and 
$-\xi+2\,\cos\,\theta_{3,4} > 0$.
The minimum condition is fulfilled in two cases:
\[
\theta_2  \platz\text{for\; all\; valid }\platz\xi,
\]
\[
\theta_3  \platz\text{for}\platz \xi<\frac12.
\]
The minimum for domain I comes from $\theta_2$  and is
\begin{eqnarray*}
u^I(\xi)&=&\sqrt{\frac12\,(1-\xi^2)
+\sqrt{\frac12\,\xi^2 \, \left(1+\frac12\, \xi^2 \right) }}.
\end{eqnarray*}
The minimum for domain  II comes from $\theta_3$ and is
\[u^{II}(\xi)=
- \sqrt{\frac12\,(1-\xi^2)
-\sqrt{\frac12\,\xi^2 \, \left(1+\frac12\, \xi^2 \right) }}.\]
The minimum for domain~II exists only if $\xi<1/2$.
We can also reformulate the reduced potential in term of $u$. It has the
form
\[\tilde{\phi}
=-\xi\, u  - \frac{1}{2}\,\sqrt{u^2-u^4} + \frac14.\]
Since $\cos\,\theta=\pm\,\sqrt{1-u^ 2}$ 
is ambiguous,
we take for the reduced potential the smaller branch of 
\[
-\frac14\,\sin\,2\theta=\pm\,\frac12\,\sqrt{u^2-u^4}.
\]
This form of the reduced potential 
is not analytic for $u=0$, but it provides the minima in the correct way.



\begin{thebibliography}{22}%
\makeatletter
\providecommand \@ifxundefined [1]{%
 \@ifx{#1\undefined}
}%
\providecommand \@ifnum [1]{%
 \ifnum #1\expandafter \@firstoftwo
 \else \expandafter \@secondoftwo
 \fi
}%
\providecommand \@ifx [1]{%
 \ifx #1\expandafter \@firstoftwo
 \else \expandafter \@secondoftwo
 \fi
}%
\providecommand \natexlab [1]{#1}%
\providecommand \enquote  [1]{``#1''}%
\providecommand \bibnamefont  [1]{#1}%
\providecommand \bibfnamefont [1]{#1}%
\providecommand \citenamefont [1]{#1}%
\providecommand \href@noop [0]{\@secondoftwo}%
\providecommand \href [0]{\begingroup \@sanitize@url \@href}%
\providecommand \@href[1]{\@@startlink{#1}\@@href}%
\providecommand \@@href[1]{\endgroup#1\@@endlink}%
\providecommand \@sanitize@url [0]{\catcode `\\12\catcode `\$12\catcode
  `\&12\catcode `\#12\catcode `\^12\catcode `\_12\catcode `\%12\relax}%
\providecommand \@@startlink[1]{}%
\providecommand \@@endlink[0]{}%
\providecommand \url  [0]{\begingroup\@sanitize@url \@url }%
\providecommand \@url [1]{\endgroup\@href {#1}{\urlprefix }}%
\providecommand \urlprefix  [0]{URL }%
\providecommand \Eprint [0]{\href }%
\providecommand \doibase [0]{http://dx.doi.org/}%
\providecommand \selectlanguage [0]{\@gobble}%
\providecommand \bibinfo  [0]{\@secondoftwo}%
\providecommand \bibfield  [0]{\@secondoftwo}%
\providecommand \translation [1]{[#1]}%
\providecommand \BibitemOpen [0]{}%
\providecommand \bibitemStop [0]{}%
\providecommand \bibitemNoStop [0]{.\EOS\space}%
\providecommand \EOS [0]{\spacefactor3000\relax}%
\providecommand \BibitemShut  [1]{\csname bibitem#1\endcsname}%
\let\auto@bib@innerbib\@empty
\bibitem [{\citenamefont {Blundell}(2011)}]{Blundell2011}%
  \BibitemOpen
  \bibfield  {author} {\bibinfo {author} {\bibfnamefont {S.}~\bibnamefont
  {Blundell}},\ }\href@noop {} {\emph {\bibinfo {title} {Magnetism in Condensed
  Matter}}}\ (\bibinfo  {publisher} {Oxford University Press, Oxford, New
  York},\ \bibinfo {year} {2011})\BibitemShut {NoStop}%
\bibitem [{\citenamefont {Wolf}(2000)}]{Wolf2000}%
  \BibitemOpen
  \bibfield  {author} {\bibinfo {author} {\bibfnamefont {W.}~\bibnamefont
  {Wolf}},\ }\href@noop {} {\bibfield  {journal} {\bibinfo  {journal} {Braz. J.
  Phys.}\ }\textbf {\bibinfo {volume} {30}},\ \bibinfo {pages} {794} (\bibinfo
  {year} {2000})}\BibitemShut {NoStop}%
\bibitem [{\citenamefont {Susuki}\ \emph {et~al.}(2004)\citenamefont {Susuki},
  \citenamefont {Gignoux}, \citenamefont {Schmitt}, \citenamefont {Shigeoka},
  \citenamefont {Canfield},\ and\ \citenamefont {Detlefs}}]{Suzuki2004}%
  \BibitemOpen
  \bibfield  {author} {\bibinfo {author} {\bibfnamefont {H.}~\bibnamefont
  {Susuki}}, \bibinfo {author} {\bibfnamefont {D.}~\bibnamefont {Gignoux}},
  \bibinfo {author} {\bibfnamefont {D.}~\bibnamefont {Schmitt}}, \bibinfo
  {author} {\bibfnamefont {T.}~\bibnamefont {Shigeoka}}, \bibinfo {author}
  {\bibfnamefont {P.}~\bibnamefont {Canfield}}, \ and\ \bibinfo {author}
  {\bibfnamefont {C.}~\bibnamefont {Detlefs}},\ }\href@noop {} {\bibfield
  {journal} {\bibinfo  {journal} {J. Magn. Magn. Mater.}\ }\textbf {\bibinfo
  {volume} {272}},\ \bibinfo {pages} {E459} (\bibinfo {year}
  {2004})}\BibitemShut {NoStop}%
\bibitem [{\citenamefont {P.C.Canfield}\ \emph {et~al.}(1997)\citenamefont
  {P.C.Canfield}, \citenamefont {S.L.Bud'ko}, \citenamefont {B.K.Cho},
  \citenamefont {A.Lacerda}, \citenamefont {D.Farrell}, \citenamefont
  {E.Johnston-Halperin}, \citenamefont {V.A.Kalatsky},\ and\ \citenamefont
  {V.L.Pokrovsky}}]{Canfield1997}%
  \BibitemOpen
  \bibfield  {author} {\bibinfo {author} {\bibnamefont {P.C.Canfield}},
  \bibinfo {author} {\bibnamefont {S.L.Bud'ko}}, \bibinfo {author}
  {\bibnamefont {B.K.Cho}}, \bibinfo {author} {\bibnamefont {A.Lacerda}},
  \bibinfo {author} {\bibnamefont {D.Farrell}}, \bibinfo {author} {\bibnamefont
  {E.Johnston-Halperin}}, \bibinfo {author} {\bibnamefont {V.A.Kalatsky}}, \
  and\ \bibinfo {author} {\bibnamefont {V.L.Pokrovsky}},\ }\href@noop {}
  {\bibfield  {journal} {\bibinfo  {journal} {Phys. Rev. B}\ }\textbf {\bibinfo
  {volume} {55}},\ \bibinfo {pages} {970} (\bibinfo {year} {1997})}\BibitemShut
  {NoStop}%
\bibitem [{\citenamefont {Amici}\ and\ \citenamefont
  {Thalmeier}(1998)}]{Amici1998}%
  \BibitemOpen
  \bibfield  {author} {\bibinfo {author} {\bibfnamefont {A.}~\bibnamefont
  {Amici}}\ and\ \bibinfo {author} {\bibfnamefont {P.}~\bibnamefont
  {Thalmeier}},\ }\href@noop {} {\bibfield  {journal} {\bibinfo  {journal}
  {Phys. Rev. B}\ }\textbf {\bibinfo {volume} {57}},\ \bibinfo {pages} {10684}
  (\bibinfo {year} {1998})}\BibitemShut {NoStop}%
\bibitem [{\citenamefont {Felner}\ and\ \citenamefont
  {Nowik}(1984)}]{Felner1984}%
  \BibitemOpen
  \bibfield  {author} {\bibinfo {author} {\bibfnamefont {I.}~\bibnamefont
  {Felner}}\ and\ \bibinfo {author} {\bibfnamefont {I.}~\bibnamefont {Nowik}},\
  }\href@noop {} {\bibfield  {journal} {\bibinfo  {journal} {J. Phys. Chem.
  Solids}\ }\textbf {\bibinfo {volume} {45}},\ \bibinfo {pages} {419} (\bibinfo
  {year} {1984})}\BibitemShut {NoStop}%
\bibitem [{\citenamefont {Czjzek}\ \emph {et~al.}(1989)\citenamefont {Czjzek},
  \citenamefont {Oestreich}, \citenamefont {Schmidt}, \citenamefont
  {{\L}atka},\ and\ \citenamefont {Tomala}}]{Czjzek1989}%
  \BibitemOpen
  \bibfield  {author} {\bibinfo {author} {\bibfnamefont {G.}~\bibnamefont
  {Czjzek}}, \bibinfo {author} {\bibfnamefont {V.}~\bibnamefont {Oestreich}},
  \bibinfo {author} {\bibfnamefont {H.}~\bibnamefont {Schmidt}}, \bibinfo
  {author} {\bibfnamefont {K.}~\bibnamefont {{\L}atka}}, \ and\ \bibinfo
  {author} {\bibfnamefont {K.}~\bibnamefont {Tomala}},\ }\href@noop {}
  {\bibfield  {journal} {\bibinfo  {journal} {J. Magn. Magn. Mater.}\ }\textbf
  {\bibinfo {volume} {79}},\ \bibinfo {pages} {42} (\bibinfo {year}
  {1989})}\BibitemShut {NoStop}%
\bibitem [{\citenamefont {Kliemt}\ and\ \citenamefont
  {Krellner}(2015)}]{Kliemt2015}%
  \BibitemOpen
  \bibfield  {author} {\bibinfo {author} {\bibfnamefont {K.}~\bibnamefont
  {Kliemt}}\ and\ \bibinfo {author} {\bibfnamefont {C.}~\bibnamefont
  {Krellner}},\ }\href@noop {} {\bibfield  {journal} {\bibinfo  {journal} {J.
  Cryst. Growth}\ }\textbf {\bibinfo {volume} {419}},\ \bibinfo {pages} {37}
  (\bibinfo {year} {2015})}\BibitemShut {NoStop}%
\bibitem [{\citenamefont {Slaski}\ \emph {et~al.}(1983)\citenamefont {Slaski},
  \citenamefont {Leciejewicz},\ and\ \citenamefont {Szytu{\l}a}}]{Slaski1983}%
  \BibitemOpen
  \bibfield  {author} {\bibinfo {author} {\bibfnamefont {M.}~\bibnamefont
  {Slaski}}, \bibinfo {author} {\bibfnamefont {J.}~\bibnamefont {Leciejewicz}},
  \ and\ \bibinfo {author} {\bibfnamefont {A.}~\bibnamefont {Szytu{\l}a}},\
  }\href@noop {} {\bibfield  {journal} {\bibinfo  {journal} {J. Magn. Magn.
  Mater.}\ }\textbf {\bibinfo {volume} {39}},\ \bibinfo {pages} {268} (\bibinfo
  {year} {1983})}\BibitemShut {NoStop}%
\bibitem [{\citenamefont {Szytu{\l}a}\ \emph {et~al.}(1984)\citenamefont
  {Szytu{\l}a}, \citenamefont {Slaski}, \citenamefont {Ptasiewicz-Bak},
  \citenamefont {Leciejewicz},\ and\ \citenamefont {A.Zygmunt}}]{Szytula1984}%
  \BibitemOpen
  \bibfield  {author} {\bibinfo {author} {\bibfnamefont {A.}~\bibnamefont
  {Szytu{\l}a}}, \bibinfo {author} {\bibfnamefont {M.}~\bibnamefont {Slaski}},
  \bibinfo {author} {\bibfnamefont {H.}~\bibnamefont {Ptasiewicz-Bak}},
  \bibinfo {author} {\bibfnamefont {J.}~\bibnamefont {Leciejewicz}}, \ and\
  \bibinfo {author} {\bibnamefont {A.Zygmunt}},\ }\href@noop {} {\bibfield
  {journal} {\bibinfo  {journal} {Solid State Commun.}\ }\textbf {\bibinfo
  {volume} {52}},\ \bibinfo {pages} {395} (\bibinfo {year} {1984})}\BibitemShut
  {NoStop}%
\bibitem [{\citenamefont {Melamud}\ \emph {et~al.}(1984)\citenamefont
  {Melamud}, \citenamefont {Pinto}, \citenamefont {Felner},\ and\ \citenamefont
  {Shaked}}]{Melamud1984}%
  \BibitemOpen
  \bibfield  {author} {\bibinfo {author} {\bibfnamefont {M.}~\bibnamefont
  {Melamud}}, \bibinfo {author} {\bibfnamefont {H.}~\bibnamefont {Pinto}},
  \bibinfo {author} {\bibfnamefont {I.}~\bibnamefont {Felner}}, \ and\ \bibinfo
  {author} {\bibfnamefont {H.}~\bibnamefont {Shaked}},\ }\href@noop {}
  {\bibfield  {journal} {\bibinfo  {journal} {J.Appl.Phys.}\ }\textbf {\bibinfo
  {volume} {55}},\ \bibinfo {pages} {2034} (\bibinfo {year}
  {1984})}\BibitemShut {NoStop}%
\bibitem [{\citenamefont {Quezel}\ \emph {et~al.}(1984)\citenamefont {Quezel},
  \citenamefont {Rossat-Mignod}, \citenamefont {Chevalier}, \citenamefont
  {Lejay},\ and\ \citenamefont {Etourneau}}]{Quezel1984}%
  \BibitemOpen
  \bibfield  {author} {\bibinfo {author} {\bibfnamefont {S.}~\bibnamefont
  {Quezel}}, \bibinfo {author} {\bibfnamefont {J.}~\bibnamefont
  {Rossat-Mignod}}, \bibinfo {author} {\bibfnamefont {B.}~\bibnamefont
  {Chevalier}}, \bibinfo {author} {\bibfnamefont {P.}~\bibnamefont {Lejay}}, \
  and\ \bibinfo {author} {\bibfnamefont {J.}~\bibnamefont {Etourneau}},\
  }\href@noop {} {\bibfield  {journal} {\bibinfo  {journal} {Solid State
  Commun.}\ }\textbf {\bibinfo {volume} {49}},\ \bibinfo {pages} {685}
  (\bibinfo {year} {1984})}\BibitemShut {NoStop}%
\bibitem [{\citenamefont {Szytu{\l}a}\ and\ \citenamefont
  {Leciejewicz}(1989)}]{Szytula1989}%
  \BibitemOpen
  \bibfield  {author} {\bibinfo {author} {\bibfnamefont {A.}~\bibnamefont
  {Szytu{\l}a}}\ and\ \bibinfo {author} {\bibfnamefont {J.}~\bibnamefont
  {Leciejewicz}},\ }\href@noop {} {\emph {\bibinfo {title} {Handbook on the
  Physics and Chemistry of Rare Earths}}},\ Vol.~\bibinfo {volume} {12}\
  (\bibinfo  {publisher} {Elsevier Sci.Publ.BV},\ \bibinfo {year} {1989})\ p.\
  \bibinfo {pages} {133}\BibitemShut {NoStop}%
\bibitem [{\citenamefont {Guettler}\ \emph {et~al.}(2016)\citenamefont
  {Guettler}, \citenamefont {Generalov}, \citenamefont {Otrokov}, \citenamefont
  {Kummer}, \citenamefont {Kliemt}, \citenamefont {Fedorov}, \citenamefont
  {Chikina}, \citenamefont {Danzenb\"acher}, \citenamefont {Schulz},
  \citenamefont {Chulkov}, \citenamefont {Koroteev}, \citenamefont
  {Caroca-Canales}, \citenamefont {Shi}, \citenamefont {Radovic}, \citenamefont
  {Geibel}, \citenamefont {Laubschat}, \citenamefont {Dudin}, \citenamefont
  {Kim}, \citenamefont {Hoesch}, \citenamefont {Krellner},\ and\ \citenamefont
  {Vyalikh}}]{Guettler2016}%
  \BibitemOpen
  \bibfield  {author} {\bibinfo {author} {\bibfnamefont {M.}~\bibnamefont
  {Guettler}}, \bibinfo {author} {\bibfnamefont {A.}~\bibnamefont {Generalov}},
  \bibinfo {author} {\bibfnamefont {M.~M.}\ \bibnamefont {Otrokov}}, \bibinfo
  {author} {\bibfnamefont {K.}~\bibnamefont {Kummer}}, \bibinfo {author}
  {\bibfnamefont {K.}~\bibnamefont {Kliemt}}, \bibinfo {author} {\bibfnamefont
  {A.}~\bibnamefont {Fedorov}}, \bibinfo {author} {\bibfnamefont
  {A.}~\bibnamefont {Chikina}}, \bibinfo {author} {\bibfnamefont
  {S.}~\bibnamefont {Danzenb\"acher}}, \bibinfo {author} {\bibfnamefont
  {S.}~\bibnamefont {Schulz}}, \bibinfo {author} {\bibfnamefont {E.~V.}\
  \bibnamefont {Chulkov}}, \bibinfo {author} {\bibfnamefont {Y.~M.}\
  \bibnamefont {Koroteev}}, \bibinfo {author} {\bibfnamefont {N.}~\bibnamefont
  {Caroca-Canales}}, \bibinfo {author} {\bibfnamefont {M.}~\bibnamefont {Shi}},
  \bibinfo {author} {\bibfnamefont {M.}~\bibnamefont {Radovic}}, \bibinfo
  {author} {\bibfnamefont {C.}~\bibnamefont {Geibel}}, \bibinfo {author}
  {\bibfnamefont {C.}~\bibnamefont {Laubschat}}, \bibinfo {author}
  {\bibfnamefont {P.}~\bibnamefont {Dudin}}, \bibinfo {author} {\bibfnamefont
  {T.~K.}\ \bibnamefont {Kim}}, \bibinfo {author} {\bibfnamefont
  {M.}~\bibnamefont {Hoesch}}, \bibinfo {author} {\bibfnamefont
  {C.}~\bibnamefont {Krellner}}, \ and\ \bibinfo {author} {\bibfnamefont
  {D.~V.}\ \bibnamefont {Vyalikh}},\ }\href@noop {} {\bibfield  {journal}
  {\bibinfo  {journal} {Sci. Rep.}\ }\textbf {\bibinfo {volume} {6}},\ \bibinfo
  {pages} {24254} (\bibinfo {year} {2016})}\BibitemShut {NoStop}%
\bibitem [{\citenamefont {Stryjewski}\ and\ \citenamefont
  {Giordano}(1977)}]{Stryjewski1977}%
  \BibitemOpen
  \bibfield  {author} {\bibinfo {author} {\bibfnamefont {E.}~\bibnamefont
  {Stryjewski}}\ and\ \bibinfo {author} {\bibfnamefont {N.}~\bibnamefont
  {Giordano}},\ }\href@noop {} {\bibfield  {journal} {\bibinfo  {journal} {Adv.
  Phys.}\ }\textbf {\bibinfo {volume} {26}},\ \bibinfo {pages} {487} (\bibinfo
  {year} {1977})}\BibitemShut {NoStop}%
\bibitem [{\citenamefont {Carra}\ \emph {et~al.}(1993)\citenamefont {Carra},
  \citenamefont {Thole}, \citenamefont {Altarelli},\ and\ \citenamefont
  {Wang}}]{Carra1993}%
  \BibitemOpen
  \bibfield  {author} {\bibinfo {author} {\bibfnamefont {P.}~\bibnamefont
  {Carra}}, \bibinfo {author} {\bibfnamefont {B.~T.}\ \bibnamefont {Thole}},
  \bibinfo {author} {\bibfnamefont {M.}~\bibnamefont {Altarelli}}, \ and\
  \bibinfo {author} {\bibfnamefont {X.}~\bibnamefont {Wang}},\ }\href@noop {}
  {\bibfield  {journal} {\bibinfo  {journal} {Phys. Rev. Lett.}\ }\textbf
  {\bibinfo {volume} {70}},\ \bibinfo {pages} {694} (\bibinfo {year}
  {1993})}\BibitemShut {NoStop}%
\bibitem [{\citenamefont {Krishnamurthy}\ \emph {et~al.}(2009)\citenamefont
  {Krishnamurthy}, \citenamefont {Keavney}, \citenamefont {Haskel},
  \citenamefont {Lang}, \citenamefont {Srajer}, \citenamefont {Sales},
  \citenamefont {Mandrus},\ and\ \citenamefont
  {Robertson}}]{Krishnamurthy2009}%
  \BibitemOpen
  \bibfield  {author} {\bibinfo {author} {\bibfnamefont {V.~V.}\ \bibnamefont
  {Krishnamurthy}}, \bibinfo {author} {\bibfnamefont {D.~J.}\ \bibnamefont
  {Keavney}}, \bibinfo {author} {\bibfnamefont {D.}~\bibnamefont {Haskel}},
  \bibinfo {author} {\bibfnamefont {J.~C.}\ \bibnamefont {Lang}}, \bibinfo
  {author} {\bibfnamefont {G.}~\bibnamefont {Srajer}}, \bibinfo {author}
  {\bibfnamefont {B.~C.}\ \bibnamefont {Sales}}, \bibinfo {author}
  {\bibfnamefont {D.~G.}\ \bibnamefont {Mandrus}}, \ and\ \bibinfo {author}
  {\bibfnamefont {J.~L.}\ \bibnamefont {Robertson}},\ }\href@noop {} {\bibfield
   {journal} {\bibinfo  {journal} {Phys. Rev. B}\ }\textbf {\bibinfo {volume}
  {79}},\ \bibinfo {pages} {014426} (\bibinfo {year} {2009})}\BibitemShut
  {NoStop}%
\bibitem [{\citenamefont {Kummer}\ \emph {et~al.}(2016)\citenamefont {Kummer},
  \citenamefont {Fondacaro}, \citenamefont {Jimenez}, \citenamefont
  {Velez-Fort}, \citenamefont {Amorese}, \citenamefont {Aspbury}, \citenamefont
  {Yakhou-Harris}, \citenamefont {van~der Linden},\ and\ \citenamefont
  {Brookes}}]{Kummer2016}%
  \BibitemOpen
  \bibfield  {author} {\bibinfo {author} {\bibfnamefont {K.}~\bibnamefont
  {Kummer}}, \bibinfo {author} {\bibfnamefont {A.}~\bibnamefont {Fondacaro}},
  \bibinfo {author} {\bibfnamefont {E.}~\bibnamefont {Jimenez}}, \bibinfo
  {author} {\bibfnamefont {E.}~\bibnamefont {Velez-Fort}}, \bibinfo {author}
  {\bibfnamefont {A.}~\bibnamefont {Amorese}}, \bibinfo {author} {\bibfnamefont
  {M.}~\bibnamefont {Aspbury}}, \bibinfo {author} {\bibfnamefont
  {F.}~\bibnamefont {Yakhou-Harris}}, \bibinfo {author} {\bibfnamefont
  {P.}~\bibnamefont {van~der Linden}}, \ and\ \bibinfo {author} {\bibfnamefont
  {N.}~\bibnamefont {Brookes}},\ }\href@noop {} {\bibfield  {journal} {\bibinfo
   {journal} {J.Synchrotron Rad.}\ }\textbf {\bibinfo {volume} {23}},\ \bibinfo
  {pages} {464} (\bibinfo {year} {2016})}\BibitemShut {NoStop}%
\bibitem [{\citenamefont {Sessi}\ \emph {et~al.}(2010)\citenamefont {Sessi},
  \citenamefont {Kuhnke}, \citenamefont {Zhang}, \citenamefont {Honolka},
  \citenamefont {Kern}, \citenamefont {Tieg}, \citenamefont {Sipr},
  \citenamefont {Minar},\ and\ \citenamefont {Ebert}}]{Sessi2010}%
  \BibitemOpen
  \bibfield  {author} {\bibinfo {author} {\bibfnamefont {V.}~\bibnamefont
  {Sessi}}, \bibinfo {author} {\bibfnamefont {K.}~\bibnamefont {Kuhnke}},
  \bibinfo {author} {\bibfnamefont {J.}~\bibnamefont {Zhang}}, \bibinfo
  {author} {\bibfnamefont {J.}~\bibnamefont {Honolka}}, \bibinfo {author}
  {\bibfnamefont {K.}~\bibnamefont {Kern}}, \bibinfo {author} {\bibfnamefont
  {C.}~\bibnamefont {Tieg}}, \bibinfo {author} {\bibfnamefont {O.}~\bibnamefont
  {Sipr}}, \bibinfo {author} {\bibfnamefont {J.}~\bibnamefont {Minar}}, \ and\
  \bibinfo {author} {\bibfnamefont {H.}~\bibnamefont {Ebert}},\ }\href@noop {}
  {\bibfield  {journal} {\bibinfo  {journal} {Phys. Rev. B}\ }\textbf {\bibinfo
  {volume} {82}},\ \bibinfo {pages} {184413} (\bibinfo {year}
  {2010})}\BibitemShut {NoStop}%
\bibitem [{\citenamefont {Ising}(1925)}]{Ising1925}%
  \BibitemOpen
  \bibfield  {author} {\bibinfo {author} {\bibfnamefont {E.}~\bibnamefont
  {Ising}},\ }\href@noop {} {\bibfield  {journal} {\bibinfo  {journal} {Z.
  Phys.}\ }\textbf {\bibinfo {volume} {31}},\ \bibinfo {pages} {253} (\bibinfo
  {year} {1925})}\BibitemShut {NoStop}%
\bibitem [{\citenamefont {Hazewinkel}(2001)}]{RAF}%
  \BibitemOpen
  \bibfield  {author} {\bibinfo {author} {\bibfnamefont {M.}~\bibnamefont
  {Hazewinkel}},\ }\href@noop {} {\emph {\bibinfo {title} {Analytic function,
  Encyclopedia of Mathematics}}}\ (\bibinfo  {publisher} {Springer,
  Heidelberg},\ \bibinfo {year} {2001})\BibitemShut {NoStop}%
\bibitem [{\citenamefont {Hill}\ and\ \citenamefont
  {McMorrow}(1996)}]{hill1996-ac}%
  \BibitemOpen
  \bibfield  {author} {\bibinfo {author} {\bibfnamefont {J.~P.}\ \bibnamefont
  {Hill}}\ and\ \bibinfo {author} {\bibfnamefont {D.~F.}\ \bibnamefont
  {McMorrow}},\ }\href@noop {} {\bibfield  {journal} {\bibinfo  {journal} {Acta
  Cryst.}\ }\textbf {\bibinfo {volume} {A52}},\ \bibinfo {pages} {236}
  (\bibinfo {year} {1996})}\BibitemShut {NoStop}%
\end{thebibliography}
%
\end{document}